
\font\bigbold=cmbx12
\font\ninerm=cmr9      \font\eightrm=cmr8    \font\sixrm=cmr6
\font\fiverm=cmr5
\font\ninebf=cmbx9     \font\eightbf=cmbx8   \font\sixbf=cmbx6
\font\fivebf=cmbx5
\font\ninei=cmmi9      \skewchar\ninei='177  \font\eighti=cmmi8
\skewchar\eighti='177  \font\sixi=cmmi6      \skewchar\sixi='177
\font\fivei=cmmi5
\font\ninesy=cmsy9     \skewchar\ninesy='60  \font\eightsy=cmsy8
\skewchar\eightsy='60  \font\sixsy=cmsy6     \skewchar\sixsy='60
\font\fivesy=cmsy5     \font\nineit=cmti9    \font\eightit=cmti8
\font\ninesl=cmsl9     \font\eightsl=cmsl8
\font\ninett=cmtt9     \font\eighttt=cmtt8
\font\tenfrak=eufm10   \font\ninefrak=eufm9  \font\eightfrak=eufm8
\font\sevenfrak=eufm7  \font\fivefrak=eufm5
\font\tenbb=msbm10     \font\ninebb=msbm9    \font\eightbb=msbm8
\font\sevenbb=msbm7    \font\fivebb=msbm5
\font\tenssf=cmss10    \font\ninessf=cmss9   \font\eightssf=cmss8
\font\tensmc=cmcsc10

\newfam\bbfam   \textfont\bbfam=\tenbb \scriptfont\bbfam=\sevenbb
\scriptscriptfont\bbfam=\fivebb  \def\Bbb{\fam\bbfam}
\newfam\frakfam  \textfont\frakfam=\tenfrak \scriptfont\frakfam=%
\sevenfrak \scriptscriptfont\frakfam=\fivefrak  \def\frak{\fam\frakfam}
\newfam\ssffam  \textfont\ssffam=\tenssf \scriptfont\ssffam=\ninessf
\scriptscriptfont\ssffam=\eightssf  
\def\smc{\tensmc}

\def\eightpoint{\textfont0=\eightrm \scriptfont0=\sixrm
\scriptscriptfont0=\fiverm  \def\rm{\fam0\eightrm}%
\textfont1=\eighti \scriptfont1=\sixi \scriptscriptfont1=\fivei
\def\oldstyle{\fam1\eighti}\textfont2=\eightsy
\scriptfont2=\sixsy \scriptscriptfont2=\fivesy
\textfont\itfam=\eightit         \def\it{\fam\itfam\eightit}%
\textfont\slfam=\eightsl         \def\sl{\fam\slfam\eightsl}%
\textfont\ttfam=\eighttt         \def\tt{\fam\ttfam\eighttt}%
\textfont\frakfam=\eightfrak     \def\frak{\fam\frakfam\eightfrak}%
\textfont\bbfam=\eightbb         \def\Bbb{\fam\bbfam\eightbb}%
\textfont\bffam=\eightbf         \scriptfont\bffam=\sixbf
\scriptscriptfont\bffam=\fivebf  \def\bf{\fam\bffam\eightbf}%
\abovedisplayskip=9pt plus 2pt minus 6pt   \belowdisplayskip=%
\abovedisplayskip  \abovedisplayshortskip=0pt plus 2pt
\belowdisplayshortskip=5pt plus2pt minus 3pt  \smallskipamount=%
2pt plus 1pt minus 1pt  \medskipamount=4pt plus 2pt minus 2pt
\bigskipamount=9pt plus4pt minus 4pt  \setbox\strutbox=%
\hbox{\vrule height 7pt depth 2pt width 0pt}%
\normalbaselineskip=9pt \normalbaselines \rm}

\def\ninepoint{\textfont0=\ninerm \scriptfont0=\sixrm
\scriptscriptfont0=\fiverm  \def\rm{\fam0\ninerm}\textfont1=\ninei
\scriptfont1=\sixi \scriptscriptfont1=\fivei \def\oldstyle%
{\fam1\ninei}\textfont2=\ninesy \scriptfont2=\sixsy
\scriptscriptfont2=\fivesy
\textfont\itfam=\nineit          \def\it{\fam\itfam\nineit}%
\textfont\slfam=\ninesl          \def\sl{\fam\slfam\ninesl}%
\textfont\ttfam=\ninett          \def\tt{\fam\ttfam\ninett}%
\textfont\frakfam=\ninefrak      \def\frak{\fam\frakfam\ninefrak}%
\textfont\bbfam=\ninebb          \def\Bbb{\fam\bbfam\ninebb}%
\textfont\bffam=\ninebf          \scriptfont\bffam=\sixbf
\scriptscriptfont\bffam=\fivebf  \def\bf{\fam\bffam\ninebf}%
\abovedisplayskip=10pt plus 2pt minus 6pt \belowdisplayskip=%
\abovedisplayskip  \abovedisplayshortskip=0pt plus 2pt
\belowdisplayshortskip=5pt plus2pt minus 3pt  \smallskipamount=%
2pt plus 1pt minus 1pt  \medskipamount=4pt plus 2pt minus 2pt
\bigskipamount=10pt plus4pt minus 4pt  \setbox\strutbox=%
\hbox{\vrule height 7pt depth 2pt width 0pt}%
\normalbaselineskip=10pt \normalbaselines \rm}

\global\newcount\secno \global\secno=0 \global\newcount\meqno
\global\meqno=1 \global\newcount\appno \global\appno=0
\newwrite\eqmac \def\romappno{\ifcase\appno\or A\or B\or C\or D\or
E\or F\or G\or H\or I\or J\or K\or L\or M\or N\or O\or P\or Q\or
R\or S\or T\or U\or V\or W\or X\or Y\or Z\fi}
\def\eqn#1{ \ifnum\secno>0 \eqno(\the\secno.\the\meqno)
\xdef#1{\the\secno.\the\meqno} \else\ifnum\appno>0
\eqno({\rm\romappno}.\the\meqno)\xdef#1{{\rm\romappno}.\the\meqno}
\else \eqno(\the\meqno)\xdef#1{\the\meqno} \fi \fi
\global\advance\meqno by1 }

\global\newcount\refno \global\refno=1 \newwrite\reffile
\newwrite\refmac \newlinechar=`\^^J \def\ref#1#2%
{\the\refno\nref#1{#2}} \def\nref#1#2{\xdef#1{\the\refno}
\ifnum\refno=1\immediate\openout\reffile=refs.tmp\fi
\immediate\write\reffile{\noexpand\item{[\noexpand#1]\ }#2\noexpand%
\nobreak.} \immediate\write\refmac{\def\noexpand#1{\the\refno}}
\global\advance\refno by1} \def\semi{;\hfil\noexpand ^^J}
\def\nl{\hfil\noexpand ^^J} \def\refn#1#2{\nref#1{#2}}

\def\ann#1#2#3{{\it Ann.\ Phys.}\ {\bf {#1}} ({#2}) #3}

\def\epl#1#2#3{{\it Europhys.\ Lett.}\ {\bf {#1}} ({#2}) #3}
\def\fp#1#2#3{{\it Found.\ Phys.}\ {\bf {#1}} ({#2}) #3}

\def\jpA#1#2#3{{\it J.\ Phys.}\ {\bf A{#1}} ({#2}) #3}

\def\plA#1#2#3{{\it Phys.\ Lett.}\ {\bf A{#1}} ({#2}) #3}
\def\pr#1#2#3{{\it Phys.\ Rev.}\ {\bf {#1}} ({#2}) #3}
\def\prA#1#2#3{{\it Phys.\ Rev.}\ {\bf A{#1}} ({#2}) #3}

\def\prl#1#2#3{{\it Phys.\ Rev.\ Lett.}\ {\bf #1} ({#2}) #3}

\def\nat#1#2#3{{\it Nature}\ {\bf {#1}} ({#2}) #3}

\newif\iftitlepage \titlepagetrue \newtoks\titlepagefoot
\titlepagefoot={\hfil} \newtoks\otherpagesfoot \otherpagesfoot=%
{\hfil\tenrm\folio\hfil} \footline={\iftitlepage\the\titlepagefoot%
\global\titlepagefalse \else\the\otherpagesfoot\fi}

\def\abstract#1{{\parindent=30pt\narrower\noindent\ninepoint\openup
2pt #1\par}}

\newcount\notenumber\notenumber=1 \def\note#1
{\unskip\footnote{$^{\the\notenumber}$} {\eightpoint\openup 1pt #1}
\global\advance\notenumber by 1}

\def\today{\ifcase\month\or January\or February\or March\or
April\or May\or June\or July\or August\or September\or October\or
November\or December\fi \space\number\day, \number\year}

\def\pagewidth#1{\hsize= #1}  \def\pageheight#1{\vsize= #1}
\def\hcorrection#1{\advance\hoffset by #1}
\def\vcorrection#1{\advance\voffset by #1}

\pageheight{23cm}
\pagewidth{15.7cm}
\hcorrection{-1mm}
\magnification= \magstep1
\parskip=5pt plus 1pt minus 1pt
\tolerance 8000
\def\bsk{\baselineskip= 15pt plus 1pt minus 1pt}
\bsk

\font\extra=cmss10 scaled \magstep0  \setbox1 = \hbox{{{\extra R}}}
\setbox2 = \hbox{{{\extra I}}}       \setbox3 = \hbox{{{\extra C}}}
\setbox4 = \hbox{{{\extra Z}}}       \setbox5 = \hbox{{{\extra N}}}




                                                        
\def\frac#1#2{{#1\over#2}}

\def\ket#1{|#1\rangle}

\def\pmb#1{\setbox0=\hbox{$#1$} \kern-.025em\copy0\kern-\wd0
    \kern.05em\copy0\kern-\wd0 \kern-.025em\raise.0433em\box0 }

\def\({\left(}
\def\){\right)}
\def\[{\left[}
\def\]{\right]}
\def\la{\langle}
\def\ra{\rangle}

\def\ca{\cos\alpha_1^\star}
\def\sa{\sin\alpha_1^\star}
\def\cb{\cos\beta_1^\star}
\def\sb{\sin\beta_1^\star}

\def\cp{\cos\phi^\star}
\def\sp{\sin\phi^\star}

\def\em{\it}

\def\SD{\epsffile{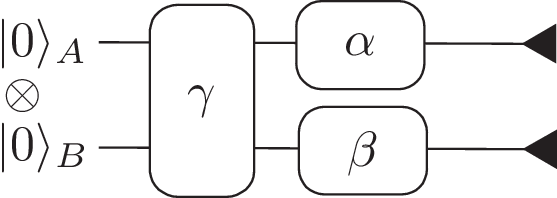}}
\def\Typeone{\epsffile{Schmidtedge.eps}}
\def\IIICatwoa{\epsffile{IIICatwo1.eps}}

\def\IIIsp{\epsffile{Schmidtsp.eps}}
\def\CGphase{\epsffile{CGphase.eps}}
\def\CGpayoff{\epsffile{CGpayoff.eps}}
\def\BoSpayoff{\epsffile{BoSpayoff.eps}}
\def\BoSphase{\epsffile{BoSphase.eps}}
\def\PDphase{\epsffile{PDphase.eps}}
\def\PDpayoffa{\epsffile{PDpayoffa.eps}}
\def\PDpayoffb{\epsffile{PDpayoffb.eps}}
\def\SHphase{\epsffile{SHphase.eps}}
\def\SHpayoff{\epsffile{SHpayoff.eps}}
\input epsf.tex


{

\refn\Patel
{N. Patel,
{\it States of play},
\nat{445}{2007}{144}}

\refn\GW
{G. Gutoski and J. Watrous,
{\it Toward a general theory of quantum games (extended abstract)},
 quant-ph/0611234}

\refn\IQzerfiv
{A. Iqbal,
{\em Studies in the theory of quantum games}, 
PhD thesis, Quaid-i-Azam University (2005),
arXiv.org, quant-ph/0503176,
{\it and rerefences therein}}

\refn\FLzerfiv
{A. Flitney,
{\em Aspects of quantum game theory}, 
PhD Thesis, Adelaide University (2005). 
{\it and rerefences therein}}%

\refn\CHzerfou
{R. Cleve, P. H{\o}yer, B. Toner and J. Watrous,
{\em Consequences and limits of nonlocal strategies},
{\it in} Proc.19th IEEE Conf. on Computational Complexity (CCC 2004)}

\refn\BBTa
{G. Brassard, A. Broadbent and A. Tapp,
{\em Quantum Pseudo-Telepathy},
\fp{35}{2005}{1877-1907}}

\refn\vNM
{J. von Neumann and O. Morgenstern,
{\em Thory of Games and Economic Behavior}, 
Princeton Univ. Press, Princeton, 1944}

\refn\HS
{J. C. Harsanyi and R. Selten,
{\it A General Theory of Equilibrium Selection in Games},
MIT Press, Cambridge, 1988}

\refn\MEninin
{D.A. Meyer,
{\em Quantum strategies},
\prl{82}{1999}{1052-1055}}

\refn\EWninnin
{J. Eisert, M. Wilkens and M. Lewenstein, 
{\em Quantum games and quantum strategies}, 
\prl{83}{1999}{3077-3080}}

\refn\NT
{A. Nawaz and A. H. Toor, 
{\em Generalized Quantization Scheme for Two-Person Non-Zero-Sum Games}, 
\jpA {\bf 37}{2004}{ 11457-11463}}

\refn\TCa
{T. Cheon,
{\em Altruistic duality in evolutionary game theory},
\plA{318}{2003}{327-332}}

\refn\TCb
{T. Cheon,
{\em Altruistic contents of quantum prisoner's dilemma},
\epl{69}{2005}{149-155}}

\refn\MW
{L. Marinatto and T. Weber,
{\em A quantum approach to static games of complete information}, 
\plA{272}{2000}{291-303}}

\refn\SOMI
{J. Shimamura, S.K. \"{O}zdemir,  F. Morikoshi and N. Imoto,
{\em Quantum and classical correlations between players in game theory}, 
\jpA{37}{2004}{4423-4436}}

\refn\BHzerone
{S.C. Benjamin and P.M. Hayden,
{\em Comments on "Quantum games and quantum strategies"}, 
\prl{87}{2001}{069801(1)}}

\refn\FH
{A. P. Flitney and L. C. L. Hollenberg,
{\em Nash equilibria in quantum games with generalized two-parameter strategies},
quant-ph/0610084, {\it Phys. Lett.} {\bf A}, in press}

\refn\LJ
{C. F. Lee and N. Johnson,
{\em  Quantum Game Theory}, 
\prA{67}{2003}{022311}}

\refn\CT
{T. Cheon and I. Tsutsui,
{\it Classical and Quantum Contents of Solvable Game Theory on Hilbert Space},
\plA{348}{2006}{147-152}}

\refn\IT
{T. Ichikawa and I. Tsutsui,
{\it Duality, Phase Structures and Dilemmas in Symmetric Quantum Games},
\ann{322}{2007}{531-551}}

\refn\EPR
{A. Einstein, B. Podolsky and N. Rosen,
{\it Can Quantum Mechanical Description of Physical Reality Be Considered Complete?},
\pr{47}{1935}{777-780}}

\refn\BEsixfou
{J.S. Bell,
{\em On the Einstein-Podolsky-Rosen paradox},
{\it Physics} {\bf 1} (1964) 195-200}

\refn\NC
{M. A. Nielsen and I. L. Chuang,
{\it Quantum Computation and Quantum Information},
Cambridge University Press, Cambridge, 2000}

\refn\AVzersev
{N. Aharon and L. Vaidman,
{\em Can quantum mechanics help to win games?}, 
arXiv:0710.1721}

\refn\Schmidt
{E. Schmidt,
{\it Zur Theorie der Linearen und Nichtlinearen 
Integralgleichungen},
{\it Math.\ Annalen.}\ {\bf 63} (1906) 433-476}

\refn\DLXSWZH
{J. Du, H. Li, X. Xu, M. Shi, J. Wu, X. Zhou and R. Han,
{\em Experimental Realization of Quantum Games on a Quantum Computer},
\prl{88}{2002}{137902}}

\refn\DLXZH
{J. Du, H. Li, X. Xu, X. Zhou and R. Han,
{\em Phase-transition-like Behavior of Quantum Games},
\jpA{36}{2003}{6551-6562}}

\refn\LL
{L. D. Landau and E. M. Lifshitz,
{\it Quantum Mechanics},
Pergamon Press, 1958}

\refn\Benjamin
{S. C. Benjamin,
{\em Comment on "A quantum approach to static games of complete information"},
\plA{277}{2000}{180-182}}

\refn\CHsixnin
{J.F. Clauser, M.A. Horn, A. Shimony and R.A. Holt,
{\em Proposed experiment to test local hidden-variable theories},
\prl{\bf49}{1969}{1804-1807}}

\refn\TSeitzer
{B.S. Tsirelson,
{\em Quantum generalizations of Bell's inequality},
{\it Lett. Math. Phys.} {\bf 4} (1980) 93-100}

%

%


%

\phantom{\rightline{\it Preliminary version}}
\vskip 1cm
\centerline{\bigbold Quantum Game Theory Based on the Schmidt Decomposition}  

\vskip 30pt

\centerline{\smc
Tsubasa Ichikawa\footnote{${}^*$}
{\eightpoint email:\quad tsubasa@post.kek.jp}
{\rm and}
Izumi Tsutsui\footnote{${}^\dagger$}
{\eightpoint email:\quad izumi.tsutsui@kek.jp}
}

\vskip 5pt

{
\baselineskip=13pt
\centerline{\it
Institute of Particle and Nuclear Studies}
\centerline{\it
High Energy Accelerator Research Organization (KEK)}
\centerline{\it
Tsukuba 305-0801, Japan}
}

\vskip 5pt
\centerline{\rm and}
\vskip 5pt

\centerline{\smc
Taksu Cheon\footnote{${}^\ddagger$}
{\eightpoint email:\quad taksu.cheon@kochi-tech.ac.jp} 
}

\vskip 3pt
{
\baselineskip=13pt
\centerline{\it Laboratory of Physics}
\centerline{\it Kochi University of Technology}
\centerline{\it Tosa Yamada, Kochi 782-8502, Japan}
}

\vskip 2.5cm
\centerline{\bf Abstract}
\abstract{We present a novel formulation of quantum game theory based on the Schmidt decomposition, which has the merit that the entanglement of quantum strategies is manifestly quantified. 
We apply this formulation to 2-player, 2-strategy symmetric games and obtain a complete set of quantum Nash equilibria.  
Apart from those available with the maximal entanglement, these quantum Nash equilibria
are extensions of the Nash equilibria in classical game theory.   The phase structure of the equilibria is determined for all values of entanglement, and thereby the possibility of resolving the dilemmas by entanglement in the game of Chicken, the Battle of the Sexes, the Prisoners' Dilemma, and the Stag Hunt,  is examined. 
We find that entanglement transforms these dilemmas with each other but cannot resolve them, except in the Stag Hunt game where the dilemma can be alleviated to a certain degree.
}   

\vfill\eject
 

\pageheight{23cm}
\pagewidth{15.7cm}
\magnification= \magstep1
\def\bsk{%
\baselineskip= 14.5pt plus 1pt minus 2pt}
\parskip=5pt plus 1pt minus 1pt
\tolerance 8000
\bsk

\overfullrule=0in


\centerline{{\bf 1. Introduction}}
\bigskip

Quantum game theory, which is a theory of games with quantum strategies, has been attracting much attention among quantum physicists and economists in recent years [\Patel, \GW] (for a review, see [\GW, \IQzerfiv, \FLzerfiv]).  
There are basically two reasons for this.
One is that quantum game theory provides a general basis to treat the quantum information processing
and quantum communication in which
plural actors try to achieve their objectives
such as the increase in communication efficiency and security [\CHzerfou, \BBTa].
The other is that it offers an extension for the existing game theory [\vNM, \HS], which is now a standard tool to analyze social behaviors among competing groups, with the prospect that newly allowed quantum strategies may overpower the conventional classical strategies, altering the known outcomes in game theory [\MEninin, \EWninnin, \NT].   An important novel element of quantum game theory is the permission of {\it correlation} between the players which are usually forbidden in the standard game theory.  
The correlation is not arbitrary:  it is furnished by {\it quantum entanglement} which is the key notion separating quantum and classical worlds, and gives a new dimension to the conventional game theory allowing us to  analyze the aspects 
(such as the altruism [\TCa, \TCb]) that are often prevalent in real situations 
and yet difficult to treat.
In other words, the quantum formulation of game theory seems to foster some sort of elements of cooperative games in the context of standard non-cooperative game theory.  As such,  it may, for instance, provide superior strategies for the players [\MEninin] or lead to resolution of the so-called dilemmas  in conventional \lq classical\rq\ game theory [\EWninnin, \MW, \SOMI].

Since the initial proposals of quantum game theory were presented, 
however, it has been recognized  [\BHzerone] that  the extended formulation suffers from several pitfalls that may nullify the advantage of quantum strategies.  One of them is 
the incompleteness problem of the strategy space, that is, that  one is just using a restricted class of strategies, rather than the entire class of strategies available in the Hilbert space of quantum states determined from the game.  As a result, different 
and, at times, conflicting conclusions have been drawn, depending on the class used in the analysis  [\FH].   
Although it is certainly possible to devise special circumstances in which only a restricted class of strategies become available, the required physical setting is untenable from operational viewpoints, because the actual implementations of these strategies do not form a closed set [\BHzerone].  Moreover, in general 
these quantum strategies do not even form a convex space, and hence one may argue that the system will not be robust against environmental disturbances.    In contrast, the Hilbert space of the entire strategies is a vector space which takes care of all dynamical changes of strategies including those generated by
the standard unitary operations in quantum mechanics and 
possible reactions from the external perturbation [\LJ].  

A game theory based on the Hilbert space is not actually difficult to construct.  Suppose that the original classical game consists of two players each of whom can choose $n$ different strategies.  In the corresponding quantum game, each of the players can resort to strategies given by quantum states belonging to the corresponding $n$-dimensional Hilbert space ${\cal H}_n$.  The total space of states, containing all possible combinations of the strategies adopted by the two players independently, ought to be given by the product space ${\cal H}_n \otimes {\cal H}_n$ which is also a Hilbert space. 

An actual scheme of realization of strategies belonging to the entire Hilbert space has been presented in [\CT] for the case $n=2$.  
There, the key concept is the trio of correlation operators
that form a dihedral algebra $D_2$, which are the building blocks of the
correlation function that generates all possible joint states in the
entire Hilbert space from  the individual states of the two players
belonging to their own Hilbert spaces.
The realization has enabled us to disentangle the classically interpretable and
purely quantum components
in the whole set of quantum strategies.  It has later been applied to all possible classes of 2-player, 2-strategy games [\IT] allowing for a full analysis of some of the games discussed typically in classical game theory.  
One of the drawbacks of this scheme, however, 
is the use of the specific operator algebra that is characteristic for 2-player, 2-strategy games: its
extension into general games will not be easy technically.

A question which has been asked persistently but has never been given a satisfactory answer in the study of quantum game theory 
is whether the quantum strategies can really be superior/advantageous than classical strategies, and if so, 
what is the physical origin of the superiority.  One may argue that,
to a large extent, the superposition of states allowed in quantum mechanics 
is analogous to classical 
probability distributions, and hence the superposition of strategies admitted in quantum game theory will be simulated by classical strategies with probability distributions, {\it i.e.}, \lq mixed strategies\rq, without yielding a substantial difference between them.   However, 
a clear distinction between the quantum superposition 
and the classical probability distribution can be found
in the nonlocal correlation of quantum strategies, which is well known since the
discovery of the EPR paradox [\EPR] and the Bell inequality [\BEsixfou].  This nonlocal
correlation of two parties, the {\it entanglement}, has become the key concept
in the development of quantum information theory [\NC], and 
it is quite natural to expect that entanglement plays a central role
in realizing the superiority of quantum strategies over the classical counterparts, if any.  In fact,  for the specific type of games adopted directly from the settings where entanglement becomes crucial, such as those used for the EPR paradox and the Bell inequality, the superiority of quantum strategies is more or less manifest  [\CHzerfou, \BBTa].   In contrast, it is also possible to find an example where the superiority is seen without entanglement [\AVzersev].   
The type of games for which we address the question of superiority of quantum strategies in the present paper is different from these:  it is the type of 2-player, 2-strategy games which are standard tools used frequently in classical game theory and have been extensively discussed in the context of quantum game theory as well.
In order to place our quantum game theory firmly in the context of quantum information theory,
it is essential to formulate the theory so 
that the role of the entanglement becomes transparent and thereby examine the outcomes of the games with respect to the entanglement attached.

That is exactly the subject we explore in this article.
Namely, we write down the strategies that span the entire Hilbert space
with the explicit use of a measure of entanglement, 
and formulate the quantum game theory based on the scheme provided there.
The important technical element in this is the use of the {\it Schmidt decomposition} [\Schmidt]
for describing joint strategies, which is available for games with two players.
{}For actual analysis of quantum games, we shall restrict ourselves to the class of 2-strategy symmetric games, which include familiar games in classical game theory such as the Chicken Game, the Battle of the Sexes (BoS), the Prisoners' Dilemma (PD) and the Stag Hunt (SH).  
We find a complete set of solutions for quantum Nash equilibria (QNE)  for the class of these games, which we classify into four types according to their game theoretical properties.  These are natural extensions of the classical Nash equilibria except for the type which arises only with the maximal entanglement and hence is genuinely quantum.  We also discuss the phase space structure of the QNE [\FH, \DLXSWZH, \DLXZH] with respect to the correlations for the joint strategies.  Using these results,  we analyze the possibility of resolving the dilemmas of the four games mentioned above.   We find that, in our scheme of quantum games, the dilemmas are somehow transformed with each other but will not be resolved under any entanglement, except for the case of SH where the dilemma is mitigated to a certain degree.  

This paper is organized as follows.  After the introductory account of quantum game theory in our formulation based on the Schmidt decomposition in Section 2, we present in Section 3 the complete set of solutions of QNE for symmetric games and thereby discuss the phase structures formed under arbitrary correlations.  We then study the problem of dilemmas and their possible resolutions for each of the four games in Section 4.  Section 5 is devoted to our conclusion and discussions.  The Appendix contains the technical detail on the QNE solutions and their classification used in the text.

\vskip1cm
\centerline
{\bf 2.  Quantum Game in the Schmidt Decomposition}
\medskip
\secno=2
\meqno=1

Our formulation of quantum game theory for 2-players follows from the idea that 
each of the players, Alice and Bob, has an individual space of strategies given by 
the respective Hilbert space ${\cal H}_A$ and ${\cal H}_B$, and that  
the {\it joint strategies} of the two players are represented by vectors (or pure states) in the total Hilbert space
${\cal H}$ given by the 
direct product ${\cal H} ={\cal H}_A \otimes {\cal H}_B$.   The two players have their own {\it payoff operators},
$A$ and $B$, which are self-adjoint operators in ${\cal H}$.   When their joint strategy is given by a vector $|\Psi\ra \in {\cal H}$, these operators provide the payoffs $\Pi_A$ and $\Pi_B$ for the respective players by the
expectation values,
$$
\Pi_A =\la\Psi |A|\Psi\ra, \qquad \Pi_B =\la\Psi |B|\Psi\ra.
\eqn\payoff
$$

To express the joint strategies systematically, we recall the
Schmidt decomposition theorem [\NC, \Schmidt] which states that any bi-partite pure state $|\Psi\ra \in {\cal H}_A\otimes{\cal H}_B$ can be expressed  in terms of some orthonormal bases $|\psi_k\ra_A\in {\cal H}_A$, $|\varphi_k\ra_B\in {\cal H}_B$ as
$$
|\Psi\ra=\sum_{k=0}^{{\rm min}[d_A-1,d_B-1]}\lambda_k\, |\psi_k\ra_A\, \vert\varphi_k\ra_B .
\eqn\Sch
$$
Here,  $\lambda_k$ are positive coefficients fulfilling $\sum_k \lambda_k^2=1$ and span over the range of the smaller one of the dimensions among 
$d_A=\dim {\cal H}_A$ and $d_B=\dim {\cal H}_B$ of the constituent Hilbert spaces.
Note that the bases $|\psi_k\ra_A$, $|\varphi_k\ra_B$ are also dependent on the state $|\Psi\ra$ under consideration.   
In furnishing a representation of the joint state, it is more convenient to use some fixed, state-independent bases $|i\ra_A$, $|j\ra_B$
for $i = 0, 1, \ldots, d_A -1$, $j = 0, 1, \ldots, d_B -1$.  
Let ${\cal U}_A(\alpha)$,  ${\cal U}_B(\beta)$ be the unitary operators relating the set of state-dependent bases and the set of state-independent bases as
$$
|\psi_i(\alpha)\ra_A={\cal U}_A(\alpha)|i\ra_A,
\qquad
|\varphi_j(\beta)\ra_B={\cal U}_B(\beta)|j\ra_B,
\eqn\rel
$$
where $\alpha$ and $\beta$ are parameters required to specify the unitary operators or, equivalently, the  state-dependent bases.
Plugging (\rel) back into the decomposition (\Sch), one realizes that 
the quantum entanglement of the joint state resides only in the coefficients $\lambda_k$, not in the parameters 
$\alpha$ and $\beta$ which are entirely specified by the local operations performed by the players.   

The foregoing observation shows that, for each of the players, all one can do is to choose the 
unitary operators ${\cal U}_A(\alpha)$,  ${\cal U}_B(\beta)$ for the change of the joint state $|\Psi\ra$, and 
for this reason, we call the local unitary operators ${\cal U}_A(\alpha)$,  ${\cal U}_B(\beta)$
as {\it local strategies} realized by the players.   It is important to recognize, 
however, that
different choice of local strategies may yield the same joint strategy $|\Psi\ra$ when combined as (\Sch).
The separation of entanglement from the local strategies which becomes available in  the Schmidt decomposition is a clear advantage of the present scheme of quantum game over other schemes which use different representations of strategy vectors in which entanglement is \lq tangled\rq\ with local operations by the individual players.  
On the other hand, a common trait of quantum game theory seen in all of these schemes is the appearance of the 
\lq external\rq\ parameters which determine the amount of entanglement.   In this respect, a quantum game provides an extension of a classical game through the introduction of the entanglement (or correlation) parameters.
Due to the independence of the entanglement from the local operations, we may interpret the third party specifying the  external parameters as a {\it referee} acting independently of the players.

{}For 2-qubit systems ($d_A=d_B=2$), one may adopt the conventional vectorial representation, $|0\ra = (1, 0)^T$ and $|1\ra = (0, 1)^T$, and introduce the shorthand notation,
$$
\left| i, j \right>= |i\ra_A |j\ra_B.
\eqn\basischoice
$$
Now, let us put $\lambda_0 = \cos{\gamma\over 2}$, $\lambda_1 = \sin{\gamma\over 2}$ with an angle parameter $0\le\gamma\le\pi$ (which is always possible by adjusting the overall sign of the state) 
for the generic 2-qubit entangled state
$|\Psi\ra$ in the Schmidt decomposition (\Sch).  We then find
$$
|\Psi(\alpha,\beta;\gamma)\ra = {\cal U}_A(\alpha)\otimes{\cal U}_B(\beta) \(\cos{\gamma\over 2} |0, 0\ra +\sin{\gamma\over 2} |1, 1\ra\).
\eqn\enpara
$$
To make the unitary operations explicit  (ignoring the phase factors which are irrelevant  in physics), we adopt the Euler angle representation  [\LL], 
$$
{\cal U}_A(\alpha) =e^{i\alpha_3 \sigma_3/2}e^{i\alpha_1 \sigma_2/2} e^{i\alpha_2 \sigma_3/2},
\qquad
{\cal U}_B(\beta) =e^{i\beta_3 \sigma_3/2}e^{i\beta_1 \sigma_2/2} e^{i\beta_2 \sigma_3/2},
\eqn\Eulb
$$
where
$ \sigma_a$, $a = 1, 2, 3$ are the Pauli matrices. The Euler angles are supposed to be in the ranges,
$$
0\le\alpha_1,\beta_1\le\pi,
\qquad
 0\le\alpha_2,\alpha_3,\beta_2,\beta_3\le2\pi.
 \eqn\area
$$ 

We remark that the parametrization (\Eulb) is degenerate with respect to the representation of strategies, that is, it does not necessarily provide a one-to-one mapping for a particular set of quantum states.   To find a more convenient expression of the quantum state and see where the mapping fails to be one-to-one, let us recombine the factors in the unitary operators as
$$
{\cal U}_A(\alpha) \otimes  {\cal U}_B(\beta)
= V_A \otimes  V_B\, e^{i(\alpha_2 + \beta_2) X/2} e^{i(\alpha_2 - \beta_2) Y/2}
\eqn\no
$$
using
$$
V_A(\alpha) = e^{i\alpha_3 \sigma_3/2}e^{i\alpha_1 \sigma_2/2}, \qquad
V_B(\beta) = e^{i\beta_3 \sigma_3/2}e^{i\beta_1 \sigma_2/2},
\eqn\no
$$
and
$$
X = {1\over 2}\left(\sigma_3 \otimes {\bf 1} +  {\bf 1} \otimes \sigma_3\right),
\qquad
Y =  {1\over 2}\left(\sigma_3 \otimes {\bf 1} -  {\bf 1} \otimes \sigma_3\right).
\eqn\no
$$
Since
$$
Y\, \ket{0, 0} = Y\, \ket{1, 1} = 0,
\eqn\no
$$
we learn that  the state  (\enpara) becomes
$$
|\Psi(\alpha,\beta;\gamma)\ra
=
V_A \otimes  V_B\, e^{i(\alpha_2 + \beta_2) X/2}\left(
\cos\frac{\gamma}{2} \,\ket{0, 0} +  \sin\frac{\gamma}{2}\, \ket{1, 1} \right).
\eqn\statea
$$
Also, using
$$
X\, \ket{0, 0} =  \ket{0, 0}, \qquad X\, \ket{1, 1} = - \ket{1, 1},
\eqn\no
$$
we observe
$$
\eqalign{
\cos\frac{\gamma}{2} \,\ket{0, 0} +  \sin\frac{\gamma}{2}\, \ket{1, 1}
&= 
e^{i{\pi\over 4}}e^{-i{\pi\over 4}X}
\left(\cos\frac{\gamma}{2} \,\ket{0, 0}   -i \sin\frac{\gamma}{2}\, \ket{1, 1} \right)
\cr
&= e^{i{\pi\over 4}} e^{-i{\pi\over 4}X} e^{i\gamma D_2/2}\,\ket{0, 0},
}
\eqn\clsub
$$
where 
$$
D_2 = \sigma_2 \otimes \sigma_2.
\eqn\no
$$

Removing the overall phase, and substituting (\clsub) back into (\statea), we finally 
arrive at the compact expression of the state,
$$
|\Psi(\alpha,\beta;\gamma)\ra
=
V_A(\alpha) \otimes  V_B(\beta) \, e^{i(\phi-\pi/2) X/2} e^{i\gamma D_2/2}\,\ket{0, 0},
\eqn\Schtwo
$$
with the phase sum,
$$
\phi = \alpha_2 + \beta_2.
\eqn\phase
$$
This expression (\Schtwo) provides a complete representation of the 2-qubit system, and furnishes a basis for our analysis of quantum game theory (see Figure 1).  
Note that this representation employs $2 + 2 = 4$ local parameters in $V_A$ and $V_B$ plus $\phi$ (which is determined by the local angles
in system $A$ and $B$ by (\phase)) as well as the entanglement angle $\gamma$.   The total number of necessary parameters is therefore 6, which is exactly the physical degrees of freedom of pure states in the 2-qubit system.  This number 6 is one degree less than the number of parameters used in (\enpara) with (\Eulb), which implies that we have an intrinsic degeneracy in describing the state by means of local operations, as seen in the combination of the phase sum (\phase).  This is the first source of the degeneracy of the representation (\enpara), which should be taken care of when we discuss the choice of strategies of the players.

\topinsert   
\centerline
{
\epsfxsize=2.2in
\epsfysize=0.8in
\SD
}
\bigskip
\abstract{{\bf Figure 1.} Our scheme of quantum game theory (see (\enpara) and (\Schtwo)). 
Starting from the initial 2-qubit joint state $\ket{0,0} = \ket{0}_A\otimes\ket{0}_B$, the referee first provides entanglement to the pure state by tuning  the parameter $\gamma$ in the Schmidt coefficients.   Knowing the value of $\gamma$, the two players, Alice and Bob, choose their local unitary operations with parameters $\alpha$ and $\beta$ in order to optimize their payoffs independently.}
\bigskip
\endinsert

To examine the content of the representation (\Schtwo), we may explicitly expand it in terms of the basis states, 
$$
\eqalign{
|\Psi(\alpha,\beta;\gamma)\ra
=\frac{1}{2}\big[
e^{i\xi_+}(\Gamma_- \cos\chi_+ +\Gamma_+ \cos\chi_-)
&\ket{0,0} -e^{-i\xi_+}(\Gamma_-\cos\chi_+  - \Gamma_+\cos\chi_-)\ket{1,1}\cr
-e^{i\xi_-}(\Gamma_-\sin\chi_+ - \Gamma_+\sin\chi_-)
&\ket{0,1}
-e^{-i\xi_-}(\Gamma_-\sin\chi_++\Gamma_+\sin\chi_-)\ket{1,0}
\big],
}
\eqn\deg
$$
where
$$
\chi_\pm = \frac{\alpha_1\pm\beta_1}{2}, \qquad \xi_\pm = \frac{\alpha_3\pm\beta_3}{2}, 
\qquad
\Gamma_\pm=\cos\frac{\gamma}{2}\pm e^{-i\phi}\sin\frac{\gamma}{2}.
\eqn\degsup
$$
We then observe that 
the representation of the state (\Schtwo) is not a one-to-one mapping 
if $\Gamma_+=0$ or $\Gamma_-=0$.  These two cases occur when  the two strategies of the players are 
maximally entangled, {\it i.e.}, when $\gamma$ takes the value,
$$
\gamma=\frac{\pi}{2},
\eqn\medeg
$$
and simultaneously the phase sum takes one of the values,
$$
\phi=p\pi,
\qquad p = 0\, \,\, \, {\rm or}\,\, \,\, 1.
\eqn\medegsup
$$
In passing, we mention that the degeneracy in the representation occurring at $p = 0$ is the one mentioned earlier [
 \Benjamin] as a source of counterstrategy.   Indeed, if Bob chooses his local strategy as the complex conjugate of Alice\rq s local strategy, that is,
 $$
{\cal U}_B(\beta(\alpha))={\cal U}_A^*(\alpha),
\eqn\commitment
$$
under $p=0$ with (\medeg), then 
the resulting state becomes
$$
|\Psi(\alpha,\beta(\alpha);\frac{\pi}{2})\ra=\frac{1}{\sqrt{2}}\,\,{\cal U}_A(\alpha)\otimes {\cal U}_A^*(\alpha)\(\ket{0,0}+\ket{1,1}\)=\frac{1}{\sqrt{2}}\(\ket{0,0}+\ket{1,1}\).
\eqn\no
$$
Thus the joint state becomes independent of the choice of the strategies adopted by the two players under the particular commitment (\commitment).   This shows that the maximally entangled case should be treated with special care in quantum game theory.

On the other hand, when we have
$\gamma = 0$, the joint strategy (\Schtwo) is decoupled (modulo a phase) into the product
$$
|\Psi(\alpha,\beta;0)\ra= |\psi(\alpha)\ra_A \cdot  |\psi(\beta)\ra_B,
\eqn\dprost
$$
of two local strategies of the players,
$$
|\psi(\alpha)\ra_A = V_A(\alpha) |0\ra_A, \qquad
|\psi(\beta)\ra_B = V_B(\beta) |0\ra_B.
\eqn\localst
$$
We observe also from  (\deg) that we can effectively work with $\phi\in[0,2\pi)$ which is half the range of the original parameters  (\area). 

In the present paper, among the generic 2-player, 2-strategy games, we are specifically interested in the cases in which the payoff operators $A$, $B$ commute with each other,
$$
[A, B] = AB - BA = 0.
\eqn\popcom
$$
We are then allowed to choose for our common basis $\{|i,j\ra \big\vert\, i,j=0,1\}$ in (\basischoice) by the basis 
which diagonalizes $A$ and $B$ simultaneously,
$$
\left< i', j' \right| A  \left| i, j \right>
=  A_{ij} \delta_{i' i}\delta_{j' j},
\qquad
\left< i', j' \right| B  \left| i, j \right>=    B_{ij} \delta_{i' i}\delta_{j' j}.  
\eqn\diag
$$
An important point is that the eigenvalues $A_{ij}$ and $B_{ij}$ can now be regarded as elements of  the payoff matrices of  a {\it classical} game if we choose the fixed bases in (\rel) as the eigenvectors of the two payoff operators in the {\it quantum} game.
Indeed, if we follow the standard interpretation of quantum mechanics that 
$$
x_i = \vert {}_A \la i |\psi(\alpha)\ra_A \vert^2, \qquad
y_j = \vert {}_B \la j |\psi(\beta)\ra_B \vert^2, 
\eqn\no
$$ 
represent the probability of Alice's strategy $|\psi(\alpha)\ra_A$ being in the state $\left| i \right>_A$ and the probability of Bob's strategy $|\psi(\beta)\ra_B$  in the state $\left| j \right>_B$, respectively, then from (\dprost) we see immediately that in the limit $\gamma = 0$ the payoffs become
$$
\Pi_A(\alpha,\beta;0) = \sum_{i, j} x_i\, A_{ij}\, y_j, \qquad 
\Pi_B(\alpha,\beta;0) = \sum_{i, j} x_i\, B_{ij}\, y_j.
\eqn\clpayoff
$$
These are precisely the payoffs of a classical game specified by the payoff matrices $A_{ij}$ and $B_{ij}$ obtained when the players resort to the {\it mixed strategies} in classical game theory by assigning probability distributions  $x_i$ and $y_j$ to their choices of strategies $(i, j)$.  This implies that  at the \lq classical limit\rq\ $\gamma = 0$
our quantum game reduces, in effect, to a classical game defined by the payoff matrices whose entries are given by the eigenvalues of the payoff operators.

To proceed further, for our later convenience we introduce the shorthand notation,
$$
\eqalign{
a_{00}&=\frac14\sum_{ij}A_{ij},
\qquad\quad\!\!\,\,\,\,\,\,
a_{03}=\frac14\sum_{ij}(-)^jA_{ij},\cr
a_{30}&=\frac14\sum_{ij}(-)^iA_{ij},
\qquad
a_{33}=\frac14\sum_{ij}(-)^{i+j}A_{ij},\cr
}
\eqn\defa
$$
and 
$$
r=\tan\frac{\gamma}{2},
\qquad
s=\frac{a_{30}}{a_{33}},
\qquad
t=\frac{a_{03}}{a_{33}}.
\eqn\rst
$$
With these shorthands (\defa),  the payoff for Alice, for example, can be concisely written as  
$$
\Pi_A(\alpha,\beta;\gamma)
=\Pi_A^{\rm pc}(\alpha,\beta;\gamma)+\Pi_A^{\rm in}(\alpha,\beta;\gamma)
\eqn\ExA
$$
with
$$
\Pi_A^{\rm pc}(\alpha,\beta;\gamma)=a_{00}+a_{33}\cos\alpha_1\cos\beta_1+\cos\gamma\,(a_{30}\cos\alpha_1 +a_{03}\cos\beta_1)
\eqn\pc
$$
and
$$
\Pi_A^{\rm in}(\alpha,\beta;\gamma)=a_{33}\sin\gamma\cos\phi\sin\alpha_1\sin\beta_1.
\eqn\inter
$$
The split of  the payoff is done here so that in the classical limit $\gamma = 0$ 
the former \lq pseudo-classical\rq\ term $\Pi_A^{\rm pc}$ survives and yields the classical payoff.  In contrast, the latter term $\Pi_A^{\rm in}$, which  is proportional to the factor $\cos\phi$, represents the 
\lq interference\rq\ effect of the local strategies which arise under nonvanishing entanglement $\gamma \ne 0$.
The split of the payoff we discussed above has also been noted in a different quantization scheme [\CT,\IT], which suggests that it is perhaps a common trait of quantum game theory which contains the classical game as a special case.

Now we are in a position to 
define the notion of equilibria (stable strategies) in quantum game theory, which is an analogue of the Nash equilibria (NE) in classical game theory. 
{}For a given $\gamma$, we call the joint strategy $(\alpha^\star,\beta^\star)$ {\it quantum Nash equilibrium (QNE)} if it satisfies
$$
\Pi_A(\alpha^\star,\beta^\star;\gamma)\ge\Pi_A(\alpha,\beta^\star;\gamma),
\qquad
\Pi_B(\alpha^\star,\beta^\star;\gamma)\ge\Pi_B(\alpha^\star,\beta;\gamma),
\eqn\qnedef
$$
for all $\alpha,\beta$. The conditions are locally equivalent to 
$$
\frac{\partial}{\partial\alpha_i}\Pi_A(\alpha,\beta^\star;\gamma)|_{\alpha=\alpha^\star}=0,
\qquad
\frac{\partial}{\partial\beta_i}\Pi_B(\alpha^\star,\beta;\gamma)|_{\beta=\beta^\star}=0,
\eqn\deriv
$$
and the convexity conditions
$$
{\cal P}_A(\alpha,\beta^\star;\gamma)|_{\alpha=\alpha^\star}\le0,
\qquad
{\cal P}_B(\alpha^\star,\beta;\gamma)|_{\beta=\beta^\star}\le0,
\eqn\negHe
$$
for the Hessian matrices,
$$
{\cal P}_A(\alpha,\beta;\gamma)_{ij}=\partial_{\alpha_i}\partial_{\alpha_j}\Pi_A(\alpha,\beta;\gamma),
\qquad
{\cal P}_B(\alpha,\beta;\gamma)_{ij}=\partial_{\beta_i}\partial_{\beta_j}\Pi_B(\alpha,\beta;\gamma).
\eqn\defHe
$$

An important class of games arise when the eigenvalues $A_{ij}$, $B_{ij}$ $(i,j=0,1)$ satisfy
$$
A_{ij}=B_{ji}.
\eqn\sym
$$
Those games with (\sym) possess an \lq symmetric\rq\ (or  \lq fair\rq) payoff assignment to the two players, 
$$
\Pi_B(\alpha,\beta;\gamma)=\Pi_A(\beta,\alpha;\gamma),
\eqn\AB
$$
which is verified from (\sym).  These are called {\it symmetric}\note{%
They are called $S$-symmetric games in [\IT] to make a distinction from the other $T$-symmetric games.} 
games and are the main subject of the present paper.

{}For symmetric quantum games, the conditions (\deriv) are simplified into
$$
\eqalign{
\sa (a_{33}\cos\beta_1^\star+a_{30}\cos\gamma) &=a_{33}\sin\gamma\cos\phi^\star\cos\alpha_1^\star\sin\beta_1^\star,\cr
\sb(a_{33}\cos\alpha_1^\star+a_{30}\cos\gamma) &=a_{33}\sin\gamma\cos\phi^\star\cos\beta_1^\star\sin\alpha_1^\star,\cr
a_{33}\sin\gamma\sin\phi^\star\sin\alpha_1^\star\sin\beta_1^\star &=0.
}
\eqn\symderiv
$$
The convexity conditions (\negHe) for Alice can be put into conditions for the eigenvalues of the Hessian matrix
${\cal P}_A(\alpha^\star,\beta^\star ;\gamma)$,
$$
\eqalign
{
\Lambda_\pm(\alpha^\star,\beta^\star;\gamma)=\frac{1}{2}&\Big\{ -\ca(a_{33}\cb+a_{30}\cos\gamma)-2a_{33}\sin\gamma\cos\phi^\star\sa\sb\cr
&\pm |\ca|\sqrt{(a_{33}\cb+a_{30}\cos\gamma)^2+(2a_{33}\sin\gamma\sin\phi^\star\sb)^2}\Big\}\le0.
}
\eqn\HessA
$$

In the following we 
seek the solutions for both (\deriv) and (\negHe) based on the expressions (\symderiv) and (\HessA). 
To exhaust all the solutions and reveal their game theoretical properties, we first consider (\symderiv) for different classes of values of $a_{30}$, $a_{33}$ and $\gamma$, and thereby find the solutions in each class, separately.  We then
reconstruct these solutions from their characteristic properties as strategies. 
The first step is presented in the Appendix, and the second step is described in Sec. 3.

Now we address the issue of degeneracy in the phase sum $\phi$, namely, that one cannot determine the respective phases 
$ \alpha_2^\star$ and $\beta_2^\star$ uniquely from the value of $\phi^*$.  
This poses an operational problem for the players, because it implies that they cannot adjust their phases $ \alpha_2^\star$ and $\beta_2^\star$ without knowing the other's choice, and  such a share of knowledge on the players' actual choices is forbidden in non-cooperative games.  As a possible resolution of this problem, 
in the present paper we assume that each player is fair-minded and determine their phases based on the 
parity division, that is, they share the same amount of phases to form the required value of $\phi^*$ by adjusting 
$$
\alpha_2^\star=\beta_2^\star = {1\over 2} \phi^*,
\eqn\fairas
$$ 
expecting the other player to do the same.  
This resolution is possible only for a phase sum, not for a phase difference, which we shall encounter later in choosing the phases $ \alpha_1^\star$ and $\beta_1^\star$  in a particular solution available under the maximal entanglement.

One can argue that the  fair-mindedness assumption is in fact consistent (or plausible) with the symmetric games we are considering.
To this end, let us consider the variations of the payoffs,
$$
\eqalign{
\delta \Pi_A(\alpha,\beta;\gamma) 
&= \Pi_A(\alpha,\beta;\gamma)|_{\phi+\delta}-\Pi_A(\alpha,\beta;\gamma)|_{\phi}, \cr
\delta \Pi_B(\alpha,\beta;\gamma) 
&= \Pi_B(\alpha,\beta;\gamma)|_{\phi+\delta}-\Pi_B(\alpha,\beta;\gamma)|_{\phi}, 
}
\eqn\vpalice
$$
under the change of the phase sum $\phi \to \phi + \delta$.   Alice will not choose the value (\fairas) if her payoff increases $\delta \Pi_A > 0$ 
at the expense of Bob's payoff $\delta \Pi_B < 0$, and  Bob will do the same if $\delta \Pi_B > 0$ while  $\delta \Pi_A < 0$.   This will not happen if
$$
\delta \Pi_A \cdot \delta \Pi_B \ge 0,
\eqn\fm
$$
which implies that the two players share a common interest as long as the variation of $\phi$ is concerned.   In that case, 
they will not wish to change the phase from the value $\phi =  \phi^*$ that optimizes the payoffs of the two players, and hence may well end up with choosing the value (\fairas).

In the general 2-player, 2-strategy games with commutative payoff operators, one finds
$$
\eqalign
{
&\delta \Pi_A(\alpha,\beta;\gamma)=a_{33}\sin\gamma\sin\alpha\sin\beta\, [\cos(\phi+\delta)-\cos(\phi)],\cr
&\delta \Pi_B(\alpha,\beta;\gamma)=b_{33}\sin\gamma\sin\alpha\sin\beta\, [\cos(\phi+\delta)-\cos(\phi)],\cr
}
\eqn\no
$$
where $b_{33} = \frac14\sum_{ij}(-)^{i+j}B_{ij}$ is defined from the payoff operator $B$ analogously to $a_{33}$ in (\defa).   
Accordingly,  the inequality (\fm) becomes 
$$
a_{33}b_{33}\ge 0.
\eqn\fmtwo
$$
The point is that, for symmetric games (\sym), one has $b_{33}=a_{33}$ and hence the inequality (\fmtwo) holds trivially, assuring the consistency of the fair-mindedness assumption we have adopted.   This observation suggests in turn that, for non-symmetric  games, the construction of quantum games requires some alternative machinery to determine the respective phases of the players, without invoking the assumption used here.

\vskip1cm
\centerline{\bf 3. Complete Set of QNE and their Phase Structures}
\bigskip
\secno=3
\meqno=1

{}For symmetric games, the conditions for QNE presented in the previous section can be handled rather easily allowing us to obtain a complete set of solutions for the conditions.  
We provide the technical detail of the procedure for reaching the solutions in the Appendix, and  here we just mention that the solutions can be classified into four types from their distinctive features as quantum strategies.
The purpose of this section is to discuss these features for each type of solutions, with a special emphasis on their phase structures, that is, the relation between the type of solutions admitted and the correlations/payoffs specifying the games.

\medskip
\noindent{{\bf 3.1. Type I solutions: pseudoclassical pure strategies}}

The first class of the solutions are given by the following four possibilities:
$$
\alpha_1^\star=k_\alpha\pi,
\qquad
\beta_1^\star=k_\beta\pi,
\qquad
k_\alpha,\, k_\beta=0,1, 
\eqn\typeone
$$
with arbitrary $\phi^\star$.
These strategies  satisfy  (\symderiv) for any symmetric (2-player,  2-strategy) quantum games under arbitrary correlations $\gamma$. 
To examine when the four possibilities in (\typeone) fulfill the convexity conditions (\HessA) and hence become QNE, we introduce
$$
H_\pm(\gamma)=a_{33}\pm a_{30}\cos\gamma,
\qquad
P_{\pm\pm\pm}(\gamma):=a_{00}\pm a_{33}\pm (a_{30}\pm a_{03})\cos\gamma .
\eqn\defP
$$
The convexity conditions (\HessA) and the expected payoffs for the type I solutions are summarized in Table I.

\topinsert
\centerline{
\vbox{
\offinterlineskip
\halign{
\strut $#$ \hfill &\quad $#$ \hfill &\quad $#$ \hfill  \cr
\noalign{\hrule}
(k_\alpha , k_\beta) & {\rm convexity \,\,conditions} & \Pi_A(\alpha^\star ,\beta^\star ; \gamma)\cr
\noalign{\hrule}
(0,0) & H_+(\gamma)\ge0 & P_{+++}(\gamma) \cr
(0,1)&  H_+(\gamma)\le0,H_-(\gamma)\le0 & P_{-+-}(\gamma)\cr
(1,0)&  H_+(\gamma)\le0,H_-(\gamma)\le0 & P_{---}(\gamma)\cr
(1,1) & H_-(\gamma)\ge0 & P_{+-+}(\gamma) \cr
\noalign{\hrule}
}
}
}
\abstract{{\bf Table I.}  The convexity conditions and Alice's payoffs $\Pi_A$ for the type I solution specified by $(k_\alpha , k_\beta)$ in (\typeone).  Bob's payoffs $\Pi_B$ are obtained from (\AB).}
\vskip 5mm
\endinsert

The situation described in Table I is depicted in Figure 2 on the \lq phase-plane\rq\ coordinated by $a_{30}\cos\gamma$ and $a_{33}$. 
On this plane, the family of correlations obtained by varying $\gamma$ for a given $A_{ij}$ ($=B_{ji}$) is shown by a horizontal line segment with the right end $(|a_{30}|, a_{33})$ and the left end $(-|a_{30}|, a_{33})$.   
One of these ends yields the classical limit $\gamma = 0$, where the solutions reduces to those corresponding to the classical pure strategy NE.  The other end $\gamma = \pi$ also yields a classical game with the payoffs $A_{\bar{i}\bar{j}}$ ($=B_{\bar{j}\bar{i}}$) which is obtained by converting the player's strategies, $\bar{i}:=1-i$, $\bar{j}:=1-j$, from the original classical game.   From the viewpoint of correlations, these provide the two extreme cases where 
the joint strategies become separable.
On the other hand,  
at the midpoint of the line $(0,a_{33})$ we have $\gamma=\pi/2$ and the strategies become maximally entangled.
On account of the fact that at the two ends the solutions become,  in effect,  classical pure NE, we 
recognize that the present QNE represent pseudoclassical pure strategies which are smoothly connected to the classical pure NE when the correlations of the individual strategies disappear.

\topinsert   
\centerline
{
\epsfxsize=2.5in
\epsfysize=2in
\Typeone
}
\vskip 5mm
\abstract{{\bf Figure 2.} Phase diagram of the type I solutions $(k_\alpha , k_\beta)$ in (\typeone).  The horizontal line segments represent four different typical families of correlations in quantum games obtained by varying $\gamma$.  One of the two ends of a line segment, indicated by a dot $\bullet$,  corresponds to the classical limit $\gamma=0$, while the midpoint $\gamma=\pi/2$ gives the maximal entanglement to the joint strategies.  The (left or right) position of the classical limit $\gamma=0$ on the line depends on the sign of $a_{30}$ of the game. }
\bigskip
\endinsert

As seen in Figure 2, the phase-plane is divided into four domains depending on the allowed combinations of the type I solutions labeled by $(k_\alpha , k_\beta)$ in (\typeone).   Observe that
these domains are \lq anti-symmetric\rq\  with respect to the $a_{33}$-axis in the sense that the interchange of the left and right domains implies the interchange $0 \leftrightarrow 1$ in the labels of the solutions $(k_\alpha , k_\beta)$.   
Since the type I solutions are pseudoclassical, to classify the properties of domains [\IT] we can use the standard classical game theoretical notions.   One of them is  the {\it Pareto optimality}, which means that  any other strategies cannot improve the payoffs of both of the two players simultaneously from those obtained by the particular strategy under consideration. 
If none of the QNE is Pareto optimal, a dilemma arises because the players would then feel that they could have 
chosen the strategy that ensures better payoffs for both of them.   Adopting the name of  the game, the Prisoners' Dilemma, which  typically suffers from this problem, we say that the dilemma is a \lq Prisoners' Dilemma (PD)\rq\ type in this paper.   Similarly, 
if the QNE is not unique, and if there is no particular reason to select one out of these QNE, then the players face a different type of dilemmas, which we call \lq Battle of the Sexes (BoS)\rq\ type, again, borrowing from the typical game possessing the same property.  
Finally, even if the QNE found in the game is unique and Pareto optimal as well, there might still be a problem if the QNE is not favorable from the viewpoint of risk.  This happens when, for instance, the QNE is {\it payoff dominant} ({\it i.e.}, it provides  the best payoffs for the players among other QNE) but not 
 {\it risk dominant} [\HS] ({\it i.e.}, it yields the best \lq average\rq$\,$payoff over the opponent's possible strategies under consideration). 
When this happens, unless the player cannot be sure about the opponent's rational behavior, there arises a dilemma of the type which we call \lq Stag Hunt (SH)\rq\ in view of the same situation observed in the SH game.   

Restricting ourselves to the type I solutions for the moment, we may consider when these dilemmas arise on the phase-plane.   Let us first examine the domain satisfying
 $$
 H_+(\gamma)\ge0,
 \qquad
 H_-(\gamma)\le0,
 \eqn\zerozero
 $$
which admits only one of the type I solutions $(k_\alpha, k_\beta)=(0,0)$.  Note that this solution is not Pareto optimal if
$$
F(\gamma)\le0,
\eqn\POi
$$
where
$$
F(\gamma):=\frac{P_{+++}(\gamma)-P_{+-+}(\gamma)}{2}=(a_{30}+a_{03})\cos\gamma
\eqn\defF
$$
measures the difference in Alice's payoff between the two strategies, $(0,0)$ and $(1,1)$. 
If (\POi) holds, one finds that the game suffers from a PD type dilemma already within the type I solutions. 

Analogously,  since the domain specified by 
  $$
 H_+(\gamma)\le0,
 \qquad
 H_-(\gamma)\ge0
 \eqn\oneone
 $$
admits only the solution $(k_\alpha, k_\beta)=(1,1)$, the Pareto optimality for this solution does not hold if
$$
F(\gamma)\ge 0.
\eqn\POii
$$
We note that the Pareto optimality of QNE is, in general, difficult to confirm because for that we need to examine the payoff for all other possible strategies, not just with QNE.

On the other hand, the domain given by
$$
H_+(\gamma)\le0,
\qquad
H_-(\gamma)\le0,
\eqn\zeroone
$$
possesses two type I solutions, $(k_\alpha, k_\beta)=(0,1)$ and $(1,0)$.  It is obvious that, if either of the two solutions is preferable for one of the players, then by the symmetry (\AB) the remaining solution is preferable for the opponent. 
Furthermore, if these two solutions are equally preferable, the player cannot choose one of them uniquely. 
Thus, these solutions come with a BoS type dilemma intrinsically.

Lastly, the domain defined by
$$
H_+(\gamma)\ge0,
\qquad
H_-(\gamma)\ge0,
\eqn\SH
$$
has two type I solutions, $(k_\alpha, k_\beta)=(0,0)$ and $(1,1)$.   Unless $P_{+++}(\gamma)=P_{+-+}(\gamma)$ is satisfied, the two players will choose the strategy which ensures a better payoff if the type I solutions are the only QNE available.  The payoff dominant solution can simultaneously be 
risk dominant (if it is measured by using the standard average) provided that
$$
G(\gamma) \ge 0,
\eqn\no
$$
where
$$
\eqalign{
G(\gamma) &:= F(\gamma) \left(\frac{P_{+++}(\gamma)+P_{-+-}(\gamma)}{2}-\frac{P_{---}(\gamma)+P_{+-+}(\gamma)}{2}\right)\cr
&=a_{30}(a_{30}+a_{03})\cos^2\gamma.
}
\eqn\nonSH
$$
The outcome of the foregoing analysis of the type I solutions is summarized in Table II.

\topinsert
\centerline{
\vbox{
\offinterlineskip
\halign{
\strut # \hfill &\quad $#$ \hfill &\quad  $#$ \hfill \cr
\noalign{\hrule}
Dilemmas & (k_\alpha,k_\beta) & {\rm Conditions}\cr
\noalign{\hrule}
PD & (0,0)&H_+(\gamma)\ge0,\, H_-(\gamma)\le0,\, F(\gamma)\ge0. \cr 
PD & (1,1)&H_+(\gamma)\le0,\, H_-(\gamma)\ge0,\, F(\gamma)\le0. \cr 
BoS & (0,1)\,{\rm and}\, (1,0)&H_+(\gamma)\le0,\, H_-(\gamma)\le0.\cr
SH &(0,0)\,{\rm and}\, (1,1)&H_+(\gamma)\ge0,\, H_-(\gamma)\ge0,\,G(\gamma)\ge0.
\cr
\noalign{\hrule}
}
}
}
\vskip 5mm
\abstract{{\bf Table II.}  Domains on the phase-plane for the type I QNE classified according to the dilemmas familiar in classical game theory.}
\vskip 5mm
\endinsert

\bigskip
\noindent{{\bf 3.2. Type II solutions:  pseudoclassical mixed strategies}}
\medskip

The type II solutions are given by
$$
\ca=\cb=s\frac{r+(-)^p}{r-(-)^p}, \qquad \phi^\star=p\pi, \qquad
p=0,1,
\eqn\TypeII
$$
with $s$ and $r$ defined in (\rst).
The convexity condition (\HessA) now reads
$$
(-)^pa_{33}\ge 0,
\eqn\HessII
$$
under which the solutions (\TypeII) are allowed for $s$ and $r$ fulfilling
$$
\left\vert s\frac{r+(-)^p}{r-(-)^p} \right\vert \le 1.
\eqn\existII
$$
Using the same phase-plane employed for the type I solutions, one can see explicitly if these type II solutions are admitted under the given payoffs and correlations (see Figure 3).   
\topinsert   
\centerline
{
\epsfxsize=2.25in
\epsfysize=2.25in
\IIICatwoa
}
\vskip 5mm
\abstract{{\bf Figure 3.} Phase diagram of the type II solutions.  The shaded area for $a_{33}\ge 0$ has the solutions $p=0$.  Each of the two shaded areas for $a_{33}\le 0$ has the solutions $p=1$.}
\bigskip
\endinsert

Under the type II solutions the players obtain the
payoffs,
$$
\Pi_A(\alpha^\star,\beta^\star;\gamma)=\Pi_B(\alpha^\star,\beta^\star;\gamma)=\frac{1}{a_{33}}\left[(a_{00}a_{33}-a_{03}a_{30})+(-)^p(a_{33}^2-a_{03}a_{30})\sin\gamma\right].
\eqn\PayTypeII
$$
In the separable limits $\gamma\rightarrow 0, \, \pi$, the second term on the right hand side of (\PayTypeII) disappears.  The players then find the payoff
 $$
\lim_{\gamma\rightarrow 0, \pi}\Pi_A(\alpha^\star,\beta^\star;\gamma)=\lim_{\gamma\rightarrow 0, \pi}\Pi_B(\alpha^\star,\beta^\star;\gamma)=\frac{A_{00}A_{11}-A_{01}A_{10}}{A_{00}-A_{01}-A_{10}+A_{11}},
\eqn\payoffmixed
$$
which is precisely the one obtained under the mixed NE of symmetric games in classical game theory.    
We also observe from (\TypeII) that  at the separable limits the condition (\existII) simplifies into $|s|\le 1$.   In view of (\defa), this condition is equivalent to
$$
A_{00}\ge A_{10}
\quad
{\rm and}
\quad
A_{11}\ge A_{01},
\quad
{\rm or}
\quad
A_{00}\le A_{10}
\quad
{\rm and}
\quad
A_{11}\le A_{01},
\eqn\existsep
$$
which are exactly the requirements for the classical mixed strategies to exist.  These results suggest that the type II solutions are actually the extended versions of the mixed strategies in quantum game theory that arise with the correlation induced by the entanglement of the individual strategies.   The effect of the correlation is seen in the second term of the payoff (\PayTypeII), which becomes maximal at the maximally entangled point $\gamma = \pi/2$ unless $a_{33}^2=a_{30}a_{03}$.   
 
\bigskip
\noindent{{\bf 3.3. Type III solutions: special strategies}}
\medskip

Let us consider the special case of the symmetric games in which we have 
$$
s=(-)^\sigma, \quad \sigma=0,\, 1,  \quad \hbox{and} \quad
a_{33}<0.
\eqn\ttreq
$$
Games whose payoff parameters obey these requirements admit  infinitely many solutions in addition to the two types of solutions discussed so far, and  
these are the Type III solutions given by the strategies satisfying
$$
\cos\gamma\ca\cb+(-)^\sigma(\ca+\cb)+\cos\gamma=0, 
\qquad
\phi^\star=\pi.
\eqn\TypeIII
$$
Note that, for a given $\gamma$, there are infinitely many combinations of $(\alpha^*_1, \beta^*_1)$ fulfilling the first condition of 
 (\TypeIII) and that they must arise symmetrically under the interchange of $\alpha^*_1$ and $\beta^*_1$.  The
 distribution of these solutions are depicted in Figure 4 on the $\ca$-$\cb$ plane.  Observe that
 the difference of the payoffs at the symmetric pair of the solutions reads
$$
\eqalign
{
\Pi_A(\alpha^\star,\beta^\star;\gamma)-\Pi_A(\beta^\star,\alpha^\star;\gamma)&=-\left[\Pi_B(\alpha^\star,\beta^\star;\gamma)-\Pi_B(\beta^\star,\alpha^\star;\gamma)\right]\cr
&=(a_{30}-a_{03})\cos\gamma\, (\ca-\cb).
}
\eqn\difpayoffsp
$$
Thus the same reasoning used in Sec. 3.1 for the BoS type dilemma applies here:  to any QNE there is a symmetric counterpart of QNE with which the players reach the dilemma of the BoS.   In
the classical limit $\gamma\rightarrow0$, the first condition of (\TypeIII) reduces to
$$
(\ca+(-)^\sigma)(\cb+(-)^\sigma)=0,
\eqn\TypwIIIsep
$$
implying that the solutions become identical to the corresponding classical NE.

\topinsert   
\centerline{
\epsfxsize=2.25in
\epsfysize=2in
\IIIsp
}
\vskip 3mm
\abstract{{\bf Figure 4.}  Given a $\gamma$, the type III solutions for $\sigma=0$ distribute along the arc determined by (\TypeIII) whose edges are $(\ca,\cb)=(-1,1)$ and $(1,-1)$ on the $\ca$-$\cb$ plane. Each number $\pi/2$, $\pi/3$, $\pi/4$, and $\pi/6$ refers to the value of $\gamma$ of its nearest upper right arc, respectively. Varying $\gamma$ sweeps out the entire square, which implies that any pair $(\alpha^*_1, \beta^*_1)$ becomes a special solution for some $\gamma$. By reflecting each arc for $\ca+\cb=0$, the arc of same $\gamma$ for $\sigma=1$ is obtained.}
\bigskip
\endinsert

Let us examine the classical game theoretical meaning of the requirements (\ttreq) assumed for the special strategies.  For the case $\sigma=0$, for instance, these requirements become
$$
A_{01}=A_{11},
\qquad
A_{10}>A_{00}.
\eqn\TypeIIIconst
$$
(The requirements for the case $\sigma=1$, on the other hand, are obtained by the conversion $(i,j)\rightarrow (\bar{i},\bar{j})$ of (\TypeIIIconst).)
Under these special payoffs (\TypeIIIconst), we find that there are indeed infinitely many mixed NE given by the probability distributions of the
strategies, 
$$
(p_A^\star, p_B^\star)=(0,x)\,\, \,  \hbox{and} \,\,\, (x,0),
\qquad
0\le x \le 1,
\eqn\clNE
$$
where $p_A$ and $p_B$ stand for the probabilities of adopting the strategy labeled by \lq 0\rq\ by 
Alice and Bob, respectively.  
The payoffs of the Nash equilibria $(p_A^\star, p_B^\star)=(0,x)$, for instance, are 
$$
\eqalign
{
\Pi_A(p_A^\star=0, p_B^\star=x)&=(1-x)A_{01}+xA_{10},\cr
\Pi_B(p_A^\star=0, p_B^\star=x)&=A_{01},
}
\eqn\clPay
$$
showing that Bob\rq s payoff degenerates infinitely for $x$.  

The payoffs of the other NE $(p_A^\star, p_B^\star)=(x, 0)$ are obtained by the interchange of $\Pi_A^\star$ and $\Pi_B^\star$.  From this one learns that 
$$
\eqalign
{
[\Pi_A(p_A^\star=0, p_B^\star=x)&-\Pi_A(p_A^\star=x', p_B^\star=0)] \cr
\times [ &\Pi_B(p_A^\star=0, p_B^\star=x)-\Pi_B(p_A^\star=x', p_B^\star=0)]\le 0,
}
\eqn\negative
$$
for $x,x'\in [0,1]$. The equality holds if either of 
$$
x=0,
\qquad
x'=0,
\qquad
A_{01}=A_{10},
\eqn\zero
$$
is satisfied.  The inequality (\negative) implies that the special classical games fulfilling (\TypeIIIconst) do
have the BoS type dilemma as their quantum extensions do.

\bigskip
\noindent{{\bf 3.4. Type IV solutions: singular strategies}}
\medskip

If the entanglement is maximal $\gamma=\pi/2$, then irrespective of the payoffs of the game, we have
two distinct solutions for QNE, one of which is given by
$$
\eqalign{
\alpha_1^\star+\beta_1^\star=\pi,
\qquad
\phi^\star&=\pi,}
\eqn\TypeIV
$$
for which the convexity condition reads 
$a_{33}<0$.  
The payoffs realized by this singular solution are
$$
\Pi_A(\alpha^\star ,\beta^\star;\gamma)=\Pi_B(\alpha^\star ,\beta^\star;\gamma) 
=a_{00}
-a_{33}
=
\frac{A_{01}+A_{10}}{2}.
\eqn\payoffIV
$$
The payoffs for the players suggest that this solution is effectively equal to the classical mixed strategies realizing the pure strategy $(0,1)$ and its conversion $(1,0)$ with equal probabilities.  This can be seen explicitly by observing that for the solution (\TypeIV)  the quantum joint state (\deg) reads
$$
|\Psi(\alpha^\star,\beta^\star;\frac{\pi}{2})\ra
= - \frac{1}{\sqrt{2}}\left\{
e^{i\xi_-}\ket{0,1}
+e^{-i\xi_-}\ket{1,0}
\right\},
\eqn\entne
$$
consisting precisely of the two states $\vert 01\ra$ and $\vert 10\ra$.

Because of the degeneracy occurring at (\TypeIV) (see (\medeg) and (\medegsup)),
this solution poses the same operational problem as the one encountered earlier, {\it i.e.},  the players find it difficult to determine the phases $\alpha_1^\star$ and $\beta_1^\star$  from the value of their sum. Since the solutions (\TypeIV) pass a criterion for $\alpha_1^\star+\beta_1^\star$ similar to (\fm), we shall again adopt the same fair-mindedness assumption for all variables of the players, that is, they resolve the problem by choosing  
$$
\alpha_1^\star=\beta_1^\star=\frac{\pi}{2},
\qquad
\alpha_2^\star=\beta_2^\star=\frac{\pi}{2},
\eqn\uni
$$
expecting the equal share with the other.

Another solution admitted at $\gamma=\pi/2$ is 
$$
\alpha_1^\star-\beta_1^\star=0,
\qquad
\phi^\star=0,
\eqn\elim
$$
for $a_{33}>0$.
Again, we have the degeneracy problem, but now in a way which is worse than the previous case, because
the condition for $\alpha_1^\star$ and $\beta_1^\star$ is now difference, not  the sum, for which the fair-mindedness assumption is of no use.   It seems for us that this problem cannot be resolved on reasonable grounds as long as the two players act independently, and for this reason we abandon the solution (\elim) as a possible strategy to resolve the dilemmas in this paper.

\vskip1cm
\centerline{\bf 4. Dilemmas in the Chicken Game, BoS, PD and SH}
\medskip
\secno=4\meqno=1
 
The discussion in the preceding section shows that players can have various QNE strategies to choose under a given symmetric pair of payoff operators and a correlation $\gamma$.   This leads us to the question whether or not 
the players can choose their strategy uniquely among the many QNE available.   More generally, given a game we are interested in the phase structures of the type of dilemmas appearing there, and thereby ask if it is possible to tune the correlation $\gamma$ such that the original dilemma in the classical game ($\gamma = 0$) disappears.  Below, we shall investigate this by the four typical examples of games,  the Chicken Game, BoS, PD and SH.  

Before we start our analysis, we recall an important notion which guides the players in choosing their strategies (and has been implicitly used in the preceding sections), {\it i.e.},  the {\it payoff-dominance} principle [\HS] of game theory which states that the players make their decisions in order to maximize their own payoffs.   It is thus instrumental to 
consider the payoff-difference between two QNE strategies for each of the players,  
$$
\eqalign
{
\Delta\Pi_A^{\mu ,\nu}(\gamma) &= \Pi_A(\alpha^{\mu\star},\beta^{\mu\star};\gamma)-\Pi_A(\alpha^{\nu\star},\beta^{\nu\star};\gamma), \cr
\Delta\Pi_B^{\mu ,\nu}(\gamma) &= \Pi_B(\alpha^{\mu\star},\beta^{\mu\star};\gamma)-\Pi_B(\alpha^{\nu\star},\beta^{\nu\star};\gamma),
}
\eqn\differences
$$
where $\mu, \nu$ label different QNE.  In our present case, these are one of the set $\{ {\rm I}_{(i,j)}, {\rm II}_{p},{\rm III, IV}\}$ of labels corresponding to the types of the solutions mentioned earlier.    Evaluation of the payoff-difference for all possible pairs of QNE allowed by the given payoff operators and $\gamma$ will provide a full list of QNE, and from this the phase structure of the games will be examined.
{}For instance, if there exists a single QNE which yields the best payoffs for both players, then obviously the players are happy to choose it and there does not arise a dilemma.    On the other hand, if the entanglement is maximal $\gamma=\pi/2$,
a BoS type dilemma necessarily arises because the type I, II, and IV solutions appear there with a degeneracy of the payoffs.

\bigskip
\noindent{\bf 4.1. Chicken Game}
\medskip
Let us first consider a symmetric game with payoffs satisfying
$$
A_{10}>A_{00}>A_{01}>A_{11}.
\eqn\CGA
$$
These conditions define the Chicken Game and are equivalent to
$$
a_{33}<0,
\qquad
|s|<1,
\qquad
t<-1.
\eqn\CGa
$$
From (\CGa), we find  $H_\pm(\gamma)<0$ for all $\gamma$, which is  (\zeroone) and hence there appear two type I solutions, $(0,1)$ and $(1,0)$.  The existence of the type II solution with $p=1$ is seen in Figure 3, while 
the type III solutions  are not allowed from (\CGa).  
The distribution of the QNE in the Chicken Game is illustrated in Figure 5.   
As we have seen in the previous section, as long as the type I solutions are concerned, for (\zeroone) we encounter 
a BoS type dilemma.   
This dilemma can be resolved if the payoffs (\PayTypeII) of the type II solutions are superior to those of the type I solution. 
To examine this possibility, we note that the type II solutions are invariant under the interchange of $\alpha$ and $\beta$, and that  from (\AB) it is sufficient to compare one of the type I solutions to the type II solutions. 
Thus,  for definiteness, in the following discussion we only consider $(0,1)$ for the type I solution.

\topinsert   
\centerline{
\epsfxsize=2.2in
\epsfysize=2in
\CGphase
\qquad\quad
\CGpayoff
}
\vskip 5mm
\abstract{{\bf Figure 5.} (Left) The phase diagram of the Chicken Game whose correlation family is shown by the two horizontal line segments possessing the opposite positions of the classical limit.  The two lines are contained in the region where there are two type I solutions $(0,1),(1,0)$ and the type II solutions with $p=1$. (Right) The payoffs of the three QNE solutions as functions of correlation $\gamma$. The differences of the line segment shapes refer to those of the types of the payoffs.}
\bigskip
\endinsert

The differences in the payoffs (\differences) between the type I and the type II solutions are 
$$
\eqalign
{
\Delta\Pi_A^{{\rm I}_{(0,1)}, {\rm II}_1}(\gamma) &= \Delta\Pi_B^{{\rm I}_{(1,0)}, {\rm II}_1}(\gamma)=a_{33}(s+1)(t-1)\left(r-uv^{-1}\right)\frac{r-1}{r^2+1},\cr
\Delta\Pi_B^{{\rm I}_{(0,1)}, {\rm II}_1}(\gamma) &= \Delta\Pi_A^{{\rm I}_{(1,0)}, {\rm II}_1}(\gamma)=a_{33}(s-1)(t+1)\left(r-u^{-1}v\right)\frac{r-1}{r^2+1},\cr
}
\eqn\CGdiff
$$
with
$$
u=\frac{s-1}{s+1}
\qquad
{\rm and}
\qquad
v=\frac{t-1}{t+1}.
\eqn\defxy
$$
{}From (\CGa) we observe
 $$
\Delta\Pi_A^{{\rm I}_{(0,1)}, {\rm II}_1}\cdot \Delta\Pi_A^{{\rm I}_{(1,0)}, {\rm II}_1}\le0
 \eqn\CGcoeff 
 $$
for all $r\ge0$, showing that none of the solutions is superior to the rest in their payoffs.  The payoffs which Alice receives under the QNE are shown in Figure 5, where we observe that,  except at the maximally entangled point $\gamma = \pi/2$,  Alice receives the best QNE at either one of the type I solutions $(k_\alpha, k_\beta)=(0,1), (1,0)$ depending on the sign of $\cos\gamma$.  By using the symmetry,  Bob receives the best QNE at the other of the type I solution.   Since the best solutions for Alice and Bob are different, we conclude that the dilemma of the Chicken Game cannot be resolved for any correlation $\gamma$ even if the full set of QNE are taken into account.

\bigskip
\noindent{\bf 4.2. Battle of the Sexes}
\medskip

The  Battle of the Sexes game is defined by the payoffs with the conditions $A_{ij}=B_{\bar{j}\bar{i}}$ supplemented by
$$
A_{00}>A_{11}>A_{01}=A_{10}.
\eqn\TBoSA
$$
This game is not a symmetric game but  belongs to another type of games which has a dual structure to the symmetric games we are considering, and because of this, they can be analyzed analogously.  The trick we use for this is the duality transformation  [\IT], which interchanges the types of the game, bringing the BoS game to the corresponding symmetric version of BoS. The transformed BoS then has  the payoffs fulfilling (\sym) and
$$
\bar{A}_{10}>\bar{A}_{01}>\bar{A}_{11}=\bar{A}_{00},
\eqn\SBoSA
$$
with $\bar{A}_{ij}$ being the payoffs after the transformation for which 
the constraints read
$$
\bar{a}_{33}<0,
\qquad
0<\bar{s}<1,
\qquad
\bar{s}+\bar{t}=0.
\eqn\SBoSa
$$

\topinsert   
\centerline{
\epsfxsize=2.2in
\epsfysize=2in
\BoSphase
\qquad\quad
\BoSpayoff
}
\vskip 5mm
\abstract{{\bf Figure 6.} (Left) The phase diagram of the BoS Game whose correlation family is shown by the line segment near the bottom. One the line appear two type I solutions $(0,1),(1,0)$ and the type II solution with $p=0$. (Right) The payoffs of the solutions for various correlations $\gamma$. 
}
\bigskip
\endinsert

Since the first and second inequalities in (\SBoSa) are ensured from (\CGa), the distribution of the solutions is similar to that of the Chicken Game (see Figure 6). The type II solutions exist and the players can still employ the fair-mindedness assumption (\fairas) 
on account of the fact that the relative phase $\phi$ and $a_{33}$ are invariant under the duality transformation.   On the other hand,
if the type IV solution (\TypeIV) derives from the solution (\elim) appearing before the transformation, then it suffers from an  operational problem inherent to (\elim).   However, as far as the resolution of dilemma is concerned, we can count out
the type IV solutions altogether without affecting the analysis of  Sec. 4.1.~to conclude that  
for any $\gamma$ the type II solution cannot yield the best payoff for Alice and Bob. 
It follows that the BoS dilemma in the game cannot be resolved by furnishing correlations in the present scheme of quantum game.

\bigskip
\noindent{\bf 4.3. Prisoners' Dilemma}
\medskip

The PD game is a symmetric game which has the payoffs obeying
$$
A_{10}>A_{00}>A_{11}>A_{01}
\qquad
{\rm and}
\qquad
2A_{00}>A_{01}+A_{10}>2A_{11}.
\eqn\PDdif
$$
To analyze the game, we first note that $a_{33}>0$ implies
$$
s<-1
\qquad
{\rm and}
\qquad
s+t>2,
\eqn\PDposi
$$
while $a_{33}<0$ implies
$$
s>1
\qquad
{\rm and}
\qquad
s+t<-2.
\eqn\PDnega
$$
In both cases, at the classical limit $\gamma = 0$ we have
$$
H_+(0)=a_{33}(1+s)<0
\qquad
{\rm and}
\qquad
H_-(0)=a_{33}(1-s)>0,
\eqn\PDcl
$$
and, hence, the only allowed QNE is the type I solution $(k_\alpha, k_\beta)=(1,1)$. Besides, since
$$
F(0)=a_{30}+a_{03}=a_{33}(s+t)>0,
\eqn\notPOcl
$$
we see that the QNE is not Pareto optimal, confirming that   at the 
classical limit the PD game with (\PDdif) has a PD type dilemma as it should.

\topinsert   
\centerline{
\epsfxsize=2.2in
\epsfysize=2in
\PDphase
}
\vskip 3mm
\abstract{{\bf Figure 7.} The phase diagram of the PD game. The upper horizontal line segment represents the correlation family
for (\PDposi), while the lower one represents the family for (\PDnega).}
\bigskip
\endinsert

Now,  let us consider the generic case of correlations $\gamma$.   First, for the case (\PDposi) we find that the existence condition of the type II solution (\existII)
cannot have a solution for any $r = \tan{\gamma/2}$.   The allowed type I solutions 
vary depending on the values of $\gamma$, as seen from the correlation family which is  shown by the upper line segment in  Figure 7.   Since the payoffs of the type I solutions 
degenerate, the BoS type dilemma occurs at $r=1$. 
In more detail, for $0\le r\le u^{-\frac{1}{2}}$ with $u$ defined in (\defxy), the only QNE is given by $(k_\alpha, k_\beta)=(1,1)$.  There, (\POii) becomes
$$
F(\gamma)=a_{33}(s+t)\cos\gamma>0,
\eqn\PDphaseone
$$
implying that  the QNE is not Pareto optimal. 
{}For $u^{-\frac{1}{2}}\le r\le u^{\frac{1}{2}}$, the allowed QNE are $(k_\alpha, k_\beta)=(0,0), (1,1)$.  There, (\nonSH) becomes
$$
G(\gamma)=a_{30}^2s(s+t)\cos^2\gamma\le0,
\eqn\PDphasetwo
$$
implying that the game has a SH type dilemma except at $r=1$ ($\gamma =\pi/2$).  
{}For $u^{\frac{1}{2}}<r$, the only QNE is $(k_\alpha, k_\beta)=(0,0)$, and we can conclude by an analogous argument that this  is not Pareto optimal.  
Summarizing the above,  we learn that the dilemma is not resolved for a PD game with (\PDposi) .
 
\topinsert   
\centerline{
\epsfxsize=2.2in
\epsfysize=2in
\PDpayoffa
\qquad\quad
\PDpayoffb
}
\vskip 3mm
\abstract{{\bf Figure 8.} (Left) Payoffs of the QNE in the PD game for the case (\PDposi). (Right) Payoffs of the PD game for the case (\PDnega). The type II solutions are not Pareto optimal for any $\gamma$, since they are surpassed in the payoffs by the other strategies ($P_{-\pm-}$, or $P_{+\pm+}$ in the some regions of $\gamma$ where they are not QNE).}
\bigskip
\endinsert

In contrast, a PD game with Eq.(\PDnega) possesses a different phase structure (see the lower segment of line in Figure 7).  The type II solution labeled by $p=1$ exists for $u<r<u^{-1}$.   In addition, the type IV solutions arises at $r=1$, where a BoS type dilemma occurs
because of the 
degeneracies of the solutions.  
The phase transition of the type I solutions occurs as follows: for $0\le r<u^\frac{1}{2}$, it is $(k_\alpha, k_\beta)=(1,1)$ for which (\PDphaseone) holds, indicating that the QNE is not Pareto optimal.  
{}For $u^{\frac{1}{2}}\le r\le u^{-\frac{1}{2}}$, the QNE are $(k_\alpha, k_\beta)=(0,1)$ and $(1,0)$ under which a BoS type dilemma occurs.  For $u^{\frac{1}{2}}< r$, the QNE is $(k_\alpha, k_\beta)=(0,0)$ which is not Pareto optimal.
The above results show that the dilemma of the game can be resolved only if the payoffs of the type II solutions are superior to those of the type I solutions and if the type II solutions are Pareto optimal among all possible strategies. 

Let us study the possibility of the resolution of the dilemma first for the domain $u<r<u^\frac{1}{2}$. 
The difference of the payoffs between the type I solution $(k_\alpha, k_\beta)=(0,0)$ (which is not the QNE in this domain) and the type II solution is
$$
\Delta\Pi_A^{{\rm I}_{(0,0)}, {\rm II}_1}=\Delta\Pi_B^{{\rm I}_{(0,0)}, {\rm II}_1}=a_{33}(s-1)(t-1)(r-u^{-1})(r-v^{-1})(r^2+1)^{-1}>0.
\eqn\PDdiffone
$$
This inequality implies that the type II solution is not Pareto optimal. For $u^{\frac{1}{2}}\le r\le u^{-\frac{1}{2}}$, on the other hand, we have 
$$
 a_{33}(s+1)(t-1)>0,
  \qquad
 a_{33}(s-1)(t+1)>0
 \quad
 {\rm and}
 \quad
 u^\frac{1}{2}<u^{-1}v<1.
 \eqn\PDcoeff
 $$
It follows that $\Delta\Pi_A^{\rm I, II} \Delta\Pi_B^{\rm I, II}\le0$, that is, the dilemma still remains.   Finally, for $u^{-\frac{1}{2}}<r<u^{-1}$, we can use the discussion for the case $u<r<u^\frac{1}{2}$  to deduce that, again, the dilemma is not resolved. 
Combining the result obtained for the case (\PDposi),  we conclude that the dilemma  in the PD game cannot be resolved by quantization in our scheme.

\bigskip
\noindent{\bf 4.4. Stag Hunt }
\medskip

The SH is a symmetric game with payoffs satisfying
$$
A_{00}>A_{10}\ge A_{11}>A_{01},
\qquad
{\rm and}
\qquad
A_{10}+A_{11}>A_{00}+A_{01}.
\eqn\SHA
$$
These constraints are equivalent to
$$
a_{33}>0,
\qquad
0>s\ge-1,
\qquad
t>1.
\eqn\SHa
$$
Note first that since $H_\pm(\gamma)\ge0$ for all $\gamma$, the type I solutions $(k_\alpha,k_\beta)=(0,0), (1,1)$ coexist for all $r$. Since the inequality (\nonSH) is not satisfied (except for $r=1$), we see that, in general, these type I solutions have a SH type dilemma.   The type II solution $p=0$ is admitted for $r\ge-u$ or $-u^{-1}\ge r\ge0$.  

Let us examine the question whether the classical SH type dilemma can be resolved by quantization. 
{}For $-u^{-1}<r<1$ or $1<r<-u$, only type I solutions (which have the SH dilemma for all $r$) are admitted and the dilemma is not resolved. For $r=1$, the BoS type dilemma arises due to the degeneracies of the payoffs of the type I solutions.  This leaves only the correlations in the region,
$$
-u^{-1}\ge r\ge0.
\eqn\corsol
$$
Here, we have $P_{+++}>P_{+-+}$ and thus both players prefer the type I solution $(k_\alpha, k_\beta)=(0,0)$ to $(k_\alpha, k_\beta)=(1,1)$. 
Also, since (\PDdiffone) holds in this region, both players prefer the type I solution $(k_\alpha,k_\beta)=(0,0)$ to the type II solution. 
Hence, the QNE $(k_\alpha, k_\beta)=(0,0)$ is payoff dominant (see Figure 9). 

In order to examine the risk dominance, we introduce the effective payoff table for given $\gamma$ in Table III with
$$
\eqalign
{
Q_\pm&=a_{00}-a_{33}s\frac{r+1}{r-1}\frac{tr^2\pm2r-t}{r^2+1}\cr
R_\pm&=a_{00}-a_{33}\frac{r+1}{r-1}\frac{(s^2\mp s\pm t)r^2\mp2tr-(s^2\pm s\mp t)}{r^2+1}.
}
\eqn\defQR
$$
Assuming that Bob adopts his three classes of the equilibria strategies with equal probabilities,
the average payoff given to Alice for the choice $k_\alpha=0$ is
$$
\langle\Pi_A\rangle_{k_\alpha=0}=\frac{1}{3}(P_{+++}+P_{-+-}+Q_-).
\eqn\SHaveone
$$
Likewise, if Alice chooses $k_\alpha=1$, the average payoff she receives is
$$
\langle\Pi_A\rangle_{k_\alpha=1}=\frac{1}{3}(P_{---}+P_{+-+}+Q_+), 
\eqn\SHavetwo
$$
and if Alice chooses the type II solution, the average payoff reads
$$
\langle\Pi_A\rangle_{\rm Type\, II}=\frac{1}{3}(R_++R_-+\Pi_A).
\eqn\SHavethree
$$

\topinsert
\centerline{
\vbox{
\offinterlineskip
\halign{
\strut  # \hfill &\vrule\quad # \hfill &\quad # \hfill &\quad #\hfill\cr
\noalign{\hrule}
Strategy & Bob $k_\beta=0$ & Bob $k_\beta=1$ & Bob Type II \cr
\noalign{\hrule}
Alice $k_\alpha=0$ &$(P_{+++},P_{+++})$ & $(P_{-+-},P_{---})$ & $(Q_-,R_+)$  \cr
Alice $k_\alpha=1$& $(P_{---},P_{-+-})$ & $(P_{+-+},P_{+-+})$& $(Q_+,R_-)$\cr
Alice Type II& $(R_+,Q_-)$ & $(R_-,Q_+)$& $(\Pi_A,\Pi_B)$\cr
\noalign{\hrule}
}
}
}
\vskip 5mm
\abstract{{\bf Table III.} Effective payoffs in a symmetric game, where $Q_\pm^{st}$ and $Q_\pm^{ts}$ are 
given in (\defQR) and $\Pi_A$ and $\Pi_B$ are the payoffs of the type II solutions.}
\vskip 5mm
\endinsert

The risk dominance of the $(0,0)$ solution with respect to the other solution $(1,1)$ requires
$$
\langle\Pi_A\rangle_{k_\alpha=0}-\langle\Pi_A\rangle_{k_\alpha=1}=-\frac{4a_{33}s}{3}\frac{r+1}{r-1}\left( 1-\frac{3r}{r^2+1}\right) >0,
\eqn\SHcriterionone
$$
in addition to (\corsol).   One can see readily that this is ensured for $r$ with
$$
\frac{3-\sqrt{5}}{2}<r\le-u^{-1},
\eqn\finalcor
$$
which requires $-\frac{1}{\sqrt{5}}<s<0$. 
Indeed, combining (\SHa) and (\corsol), one finds that (\SHcriterionone) turns into 
$\frac{3-\sqrt{5}}{2}<r<\frac{3+\sqrt{5}}{2}$ and hence, if $\frac{3-\sqrt{5}}{2}<-u^{-1}$ (or $-\frac{1}{\sqrt{5}}<s<0$), the inequality (\finalcor) never holds.  Conversely, (\finalcor) satisfies (\SHcriterionone) and (\corsol), which guarantees the existence of the type II solution.

On the other hand, the risk dominance of the $(0,0)$ solution with respect to the type II solutions demands
 $$
 \langle\Pi_A\rangle_{k_\alpha=0}-\langle\Pi_A\rangle_{\rm Type II}=\frac{2a_{33}(s-1)}{3}\frac{(r+u^{-1})(sr^2+2-s)}{(r^2+1)(r-1)}>0,
 \eqn\SHcriteriontwo
 $$
in addition to (\corsol). 
This, however, cannot be fulfilled, as one can see by using an argument analogous  to the one used above.  

To summarize, in the classical limit $\gamma = 0$ the average payoffs have the relations,
$$
\langle\Pi_A\rangle_{k_\alpha=0}<\langle\Pi_A\rangle_{k_\alpha=1},
\qquad
\langle\Pi_A\rangle_{k_\alpha=0}<\langle\Pi_A\rangle_{\rm Type II},
\eqn\ordercl
$$
while, under appropriate correlations, we can have
$$
\langle\Pi_A\rangle_{k_\alpha=1}<\langle\Pi_A\rangle_{k_\alpha=0}<\langle\Pi_A\rangle_{\rm Type II}.
\eqn\ordernew
$$
We thus reach the
conclusion that, although the dilemma in the SH game cannot be resolved completely, it can be weakened by alleviating the situation to a certain extent.   The analysis in the region $r>-u$ can be made similarly, yielding a similar conclusion.  

\topinsert   
\centerline{
\epsfxsize=2.2in
\epsfysize=2in
\SHphase
\qquad\quad
\SHpayoff
}
\vskip 5mm
\abstract{{\bf Figure 9.} (Left) The phase diagram of the SH game.  The horizontal line segment in the upper part represents the correlation family of the game.  (Right) The payoffs of the QNE in the family.}
\bigskip
\endinsert

Alternatively, for the examination of the risk dominance, one may consider the averages of payoffs taken over all possible quantum strategies of the opponent, that is,
 $$
 \eqalign
 {
 \langle\Pi_A\rangle_{k_\alpha=0}=&\frac{\int_0^{2\pi}d\beta_2\int_0^{\pi}d\beta_1\Pi_A(\alpha^{\rm I\star},\beta ;\gamma)}{\int_0^{2\pi}d\beta_2\int_0^{\pi}d\beta_1},\cr
 \langle\Pi_A\rangle_{k_\alpha=1}=&\frac{\int_0^{2\pi}d\beta_2\int_0^{\pi}d\beta_1\Pi_A(\alpha^{\rm I\star},\beta ;\gamma)}{\int_0^{2\pi}d\beta_2\int_0^{\pi}d\beta_1},\cr
 \langle\Pi_A\rangle_{\rm Type\, II}=&\frac{\int_0^{2\pi}d\beta_2\int_0^{\pi}d\beta_1\Pi_A(\alpha^{\rm II\star},\beta ;\gamma)}{\int_0^{2\pi}d\beta_2\int_0^{\pi}d\beta_1}.\cr
 }
 \eqn\defave
 $$
 This time, the payoff differences become simpler:
 $$
 \eqalign{
 \langle\Pi_A\rangle_{k_\alpha=0}-\langle\Pi_A\rangle_{k_\alpha=1} &=2a_{33}s\frac{1-r^2}{1+r^2}, \cr
  \langle\Pi_A\rangle_{k_\alpha=0}-\langle\Pi_A\rangle_{\rm Type II}  &=a_{33}s(s-1)\frac{(r+1)(r+u^{-1})}{r^2+1}.
  }
  \eqn\redeftwo
$$
However, from (\SHa) we learn that the ordering in the average payoffs for the correlations (\corsol) remains unchanged from the classical case (\ordercl).

\vskip1cm
\centerline{\bf 5. Conclusion and Discussions}
\medskip
\secno=5\meqno=1

In this article, we have presented a new formulation 
of quantum game theory  for 2-players.  In particular, we have provided a salutary scheme for symmetric games with 2-strategies and thereby analyzed the outcomes of the games in detail.
Our formulation is based on the Schmidt decomposition of two partite quantum states, which is an alternative to the one proposed recently in [\CT] and is 
readily extendable to $n$-strategy games.   Technically, the difference between the two formulations lies in the operator ordering of
correlation and individual local unitary transformations required to specify the joint strategy. 

As in the previous formulation, 
the present formulation is intended to accommodate all possible strategies realized in the Hilbert space (which is the state space of quantum theory)
to remedy the defect found in many of the formulations of quantum game theory proposed earlier.  
In our scheme of  quantizing a classical game, we have the 
set of correlation parameters  $\gamma$ which are determined independently of the strategies of the players.   Since the limit $\gamma \to 0$ restores the original classical game, we see that $\gamma$ provides an extension of a classical game yielding a  $\gamma$-deformed family of games.   On account of the independence of the parameters $\gamma$ from the players choice, one may think of a third party, or a  \lq referee\rq,  who tunes $\gamma$ in the game theoretic settings.    It is worth mentioning that in our scheme of quantum game theory the payoff splits into a pseudo-classical component and the rest, such that the former amounts to the payoff of the  $\gamma$-deformed family of games while the latter to the extra factor allowed only under the presence of interference and correlation.    

The present formulation turns out to be quite convenient also in analyzing the QNE, that is, the stable strategies the players would choose in quantum game theory.   Indeed, we are able to find a complete set of solutions for the equilibria, which are classified into four types in the text,
among which three are $\gamma$-deformed versions of classical
Nash equilibria, and the other one is admitted only with the maximal entanglement and hence cannot be found in classical games.  
Besides the elements that determine the original classical game from which the quantum game is defined, the existence of these equilibria depends strongly on the correlations $\gamma$ given.  The analysis of the dependence has allowed us to obtain a clear picture of the phase structure of the QNE in the game, which can be convoluted when some of the four types of solutions coexist.   We mention that
the phase structure we found shares some properties similar to those obtained by other schemes [\FH, \IT]. Since our scheme deals with the whole Hilbert space, one may argue that this similarity comes as a partial manifestation of the full phase structure obtained by our scheme. 

One of the  interests in game theory lies in learning the mechanism underlying the appearance of dilemmas and their possible resolutions.
In this respect, we need to address the question if the quantization ({\it i.e.}, the extension by introducing quantum correlations) of a classical game can provide a resolution of the dilemma, which has actually been the main thrust in the investigation of quantum game theory since its inception [\MEninin, \EWninnin, \MW].   To find the answer in our scheme of quantum game, we have investigated the
four examples of 2-player, 2-stratey games, the Chicken Game, the
(S-symmetric version of) Battle of the Sexes, the
Prisoners' Dilemma, and the Stug Hunt, all of which are plagued with dilemmas.
The outcome is somewhat discouraging, however.  Namely, we have seen that 
the players of none of the four games find a resolution of the dilemma, except for the
Stug Hunt game where the dilemma can be mitigated to some extent within the analysis done with the assumptions made there.   We note that these results are obtained with the full set of QNE, which now include the new types of equilibria absent in classical game theory.   Although this does not exclude the possibility of resolution of dilemmas in other games by quantization, it certainly suggests the generic difficulty which will arise in quantum game theory formulated in our scheme.

These results in the resolutions of the dilemmas are clearly different from those found in other literatures [\EWninnin, \TCb, \MW, \SOMI, \CT, \IT].  This originates in the differences in the quantization scheme, that is, in the treatment of local strategies of the players and  in the presence/absence of (artificial) restrictions of the state space.  For example, consider the quantization of the Battle of the Sexes, where the appearance of coexisting QNE leads to the typical dilemma for any values of correlation $\gamma$. 
Assume that a restriction of the state space does not affect the dilemma at the classical $\gamma=0$ limit (to the authors' knowledge,  all  the quantization schemes proposed so far adopt this assumption).   If one can devise a restriction such that it removes one of the coexisting QNE for some $\gamma$ and simultaneously ensures the criterion of the Pareto optimality for the remaining QNE in the restricted state space, then the game will no longer suffer from any dilemma at that value of $\gamma$.  This is in fact a standard mechanism of resolving the dilemma in different quantization schemes at the cost of introducing artificial restrictions rendering the space of strategies 
untenable from operational viewpoints  [\BHzerone]. 

{}Finally, we mention the obvious merit of the Schmidt decomposition in our formulation, that is, that the  
correlation between the strategies of the individual players is expressed in terms of a variable that 
directly specifies the degree of quantum entanglement.   This implies that the quantum 
game theory in our formulation is ready to be positioned properly in the field of quantum information.
In fact, it is possible to extend the scope of the games by considering, for instance, the cases where the two payoff operators do not commute ({\it i.e.} by eliminating our assumption (\popcom)), or the cases where the game consists of 
multiple rounds of subgames with different payoff operators.  
These yield a setup of games similar to the one given by the CHSH game [\CHsixnin, \TSeitzer],
where the quantum nonlocality will be seen to affect the outcome of the game directly.  
This would become a focus of attention in future researches, along with the
technical extension of the theory beyond the class of quantum games considered here.

\bigskip
\noindent
{\bf Acknowledgement:}
This work is supported by the Grant-in-Aid for Scientific Research, 
Nos.~13135206, 16540354 and 18540384 of the Japanese Ministry of Education, 
Science, Sports and Culture.

\vfill\eject
\centerline{\bf Appendix. Quantum Nash Equilibria and their Classification}
\medskip
 \secno=0\appno=1\meqno=1
 
In this appendix, we provide a complete set of solutions for the extremal condition (\symderiv) and the convexity condition (\HessA) for QNE in symmetric quantum games.  
Our strategy to find the complete set is as follows: first, we classify the symmetric games into four types (I) - (IV) depending on whether  $a_{33}$ and/or $a_{30}$ vanish or not.  Secondly, in each class, we obtain the solutions for the conditions separately for the separable states ($\gamma=0, \, \pi$),  the generic (nonmaximal) entangled states, and the maximally entangled states ($\gamma=\pi/2$).  The criterion for the finer classification in each class of the states derives from the different ways to meet the third equation in (\symderiv), and this procedure exhausts all possibilities for the strategies to become QNE.   Finally, we regroup the solutions to four new types which are convenient for our discussions in the text.

\vskip 5mm
\centerline{\bf{(I) $a_{33}=0$ and $a_{30}=0$}}
\medskip

In this case, both the conditions (\symderiv) and (\HessA) are trivially satisfied, 
and any set of values of $\alpha_1$, $\beta_1$, and $\phi$ provides a QNE for all correlations $\gamma$. 

\vskip 5mm
\centerline{\bf{(II) $a_{33}=0$ and $a_{30}\neq0$}}
\medskip

Eq.(\symderiv) is then simplified as
$$
\cos\gamma\sin\alpha_1^\star=\cos\gamma\sin\beta_1^\star=0.
\eqn\excep
$$
When the state is separable, $\gamma=c\pi$, $c=0,1$, we have the solutions
$$
(\alpha_1^\star,\beta_1^\star)=(k_\alpha,k_\beta)\pi,
\qquad
k_\alpha,k_\beta=0,1.
\eqn\pure
$$
Eq.(\negHe) shows that the solutions for (\symderiv) are provided by $(k_\alpha, k_\beta, c)=(0,0,0), (1,1,1)$ for $a_{30}>0$ and $(k_\alpha, k_\beta, c)=(1,1,0), (0,0,1)$ for $a_{30}<0$.
The convexity condition (\HessA) is then examined to find that 
$(k_\alpha,k_\beta)=(0,0)$ becomes QNE for $a_{30}\cos\gamma>0$ and $(k_\alpha,k_\beta)=(1,1)$ becomes QNE for $a_{30}\cos\gamma<0$.
At $\gamma=\pi/2$, both the extremal condition and the convexity conditions are fulfilled trivially, and hence any set of values of $\alpha_1$, $\beta_1$, and $\phi$ gives a QNE.

\vskip 5mm
\centerline{\bf{(III) $a_{33}\neq0$ and $a_{30}=0$}}
\medskip

\medskip
\noindent
{\rm (III-1) Separable states:}

Eq.(\symderiv) becomes 
$$
\sin\alpha_1^\star\cos\beta_1^\star=\cos\alpha_1^\star\sin\beta_1^\star=0,
\eqn\IIisep
$$
allowing for two classes of solutions.  One is given by (\pure) for which the convexity condition reads
$$
(-)^{k_\alpha+k_\beta}a_{33}\ge0.
\eqn\HeIIi
$$
The other is $\alpha_1^\star=\beta_1^\star=\pi/2$ for which no further condition arises from the convexity condition.

\medskip
\noindent
{\rm (III-2) Generic entangled states:}

Eq.(\symderiv) becomes 
$$
\eqalign{
&\sa\cb=\sin\gamma\cp\ca\sb,\cr
&\ca\sb=\sin\gamma\cp\sa\cb,\cr
&\sp\sin\alpha_1^\star\sin\beta_1^\star=0.
}
\eqn\IIia
$$
(i) For $\phi^\star=p\pi$ with $p=0,1$, there are two classes of solutions. One is (\pure) for which the convexity condition is given by (\HeIIi). The other is $\alpha_1^\star=\beta_1^\star=\pi/2$ for which the convexity condition is $(-)^pa_{33}\ge0$.

\noindent
(ii) For $\phi\neq p\pi$,  the solutions are given by (\pure) under the convexity condition (\HeIIi).

\medskip
\noindent
{\rm (III-3)  Maximally entangled states:}

Eq.(\symderiv) becomes
$$
\eqalign{
\cp\cos\alpha_1^\star\sin\beta_1^\star&=\sin\alpha_1^\star\cos\beta_1^\star,\cr
\cp\cos\beta_1^\star\sin\alpha_1^\star&=\sin\beta_1^\star\cos\alpha_1^\star,\cr
\sp\sin\alpha_1^\star\sin\beta_1^\star&=0.}
\eqn\MESIIi
$$
(i) For  $\phi^\star=p\pi$, the first and second equations of (\MESIIi) reduce to
$$
\sin(\alpha_1^\star-(-)^p\beta_1^\star)=0,
\eqn\MESIIia
$$
which provides the solutions,
$$
\alpha_1^\star=(-)^p\beta_1^\star+q\,\pi,
\qquad
q=0,1,
\eqn\solIIia
$$
where the values of $q$ are required to obey $(-)^qa_{33}>0$ by the convexity condition.  From the range of the parameters $\alpha_1^\star, \beta_1^\star$, the combinations of $(p,q)$ are restricted to $(p,q)=(0,0),(1,1)$.

\noindent
(ii) For $\phi\neq p\pi$, (\pure) gives the solution and the convexity condition reads (\HeIIi).

\vskip 5mm
\centerline{\bf (IV) $a_{33}\neq0$ and $a_{30}\neq0$}
\medskip

\medskip
\noindent
{\rm (IV-1)  Separable states:}

Eq. (\symderiv) becomes
$$
\eqalign{
\sa\left[\cb+(-)^cs\right]&=0,\cr
\sb\left[\ca+(-)^cs\right]&=0,
}
\eqn\SepIIii
$$
and the convexity conditions are
$$
\eqalign{
a_{33}\ca\left[\cb+(-)^cs\right]&\ge0,\cr
a_{33}\cb\left[\ca+(-)^cs\right]&\ge0.
}
\eqn\HeIIiiSep
$$
(i) If $\alpha_1^\star=k_\alpha\pi$ and $s=(-)^{c+k_\alpha+1}$, (\SepIIii) is satisfied.  From the convexity conditions, we find  that if $a_{33}>0$, then $k_\alpha=0$ and $\beta_1^\star=0$, or $k_\alpha=1$ and $\beta_1^\star=\pi$. If $a_{33}<0$, on the other hand, no restriction for $\beta_1^\star$ arises.

\noindent
(ii) If $\alpha_1^\star=k_\alpha\pi$ and $s\neq(-)^{c+k_\alpha+1}$,  then $\beta_1^\star=k_\beta\pi$. These  four solutions are subject to (\HeIIiiSep).

\noindent
(iii) If $\alpha_1^\star\neq k_\alpha\pi$, from the first of (\SepIIii) we have $\cb=(-)^{c+1}s$. This requires $|s| \le1$.  The second of (\SepIIii) implies either $s=(-)^\sigma$ with $\sigma=0,1$ or $\cos\alpha_1^\star=(-)^{c+1}s$.
For the former case, the convexity conditions yield  
$(-)^{c+\sigma}=1$ and $a_{33}<0$, or $(-)^{c+\sigma}=-1$ and $a_{33}<0$. For the latter case with $s\neq(-)^\sigma$, the convexity conditions are always fulfilled.


\medskip
\noindent
{\rm (IV-2)  Generic entangled states:}

\noindent
(i) Suppose first that $\phi^\star=p\pi$. One class of solutions available is then (\pure). The convexity conditions are those given in Table I. If (\pure) is not fulfilled, then we can derive
$$
\cos\gamma\cos\alpha_1^\star\cos\beta_1^\star+s(\cos\alpha_1^\star+\cos\beta_1^\star)+s^2\cos\gamma=0,
\eqn\multip
$$
from (\symderiv) by multiplying each side of the first equation by the same side of the second equation.  Furthermore, if
$$
\cos\beta_1^\star\ne0
\qquad
{\rm and}
\qquad
\cos\alpha_1^\star+s\cos\gamma\ne0
\eqn\condIIIC
$$
holds, then each side of the first equation of (\symderiv) can be devided by the same side of the second equation, respectively, leading to
$$
(\cos\alpha_1^\star-\cos\beta_1^\star)\,(s\cos\gamma\cos\alpha_1^\star\cos\beta_1^\star+\cos\alpha_1^\star+\cos\beta_1^\star+s\cos\gamma)=0.
\eqn\devide
$$
Note that solutions  for (\multip) and (\devide) do not necessarily fulfill the first and the second equations of (\symderiv), and we need to examine if they truly become the solutions by substituting them into (\symderiv) under $\phi^\star=p\pi$. 

To proceed, we first seek solutions which satisfy $\cos\alpha_1^\star=\cos\beta_1^\star$. 
The solutions of (\multip) and (\devide) are then
$$
\ca=\cb=s\frac{r+(-)^{k_r}}{r-(-)^{k_r}},
\qquad
k_r=0,1.
\eqn\solIIICatwo
$$
By substituting this to (\symderiv), we obtain 
$
k_r=p.
$
The solutions are available when the condition $|s\frac{r+(-)^{k_r}}{r-(-)^{k_r}}|\le1$ and the convexity condition
$
(-)^pa_{33}\ge0
$
are both met.

Secondly, if $\cos\alpha_1^\star\neq\cos\beta_1^\star$, we consider the solutions separately depending on whether $s=(-)^\sigma$ or not.  If we have $s=(-)^\sigma$, then (\multip) and (\devide) are combined as
$$
\cos\gamma\ca\cb+(-)^\sigma(\ca+\cb)+\cos\gamma=0.
\eqn\IIICathree
$$
Thus, all pairs of $\alpha_1^\star$ and $\beta_1^\star$ satisfying (\IIICathree) become the solutions. 
By substituting them to (\symderiv), we obtain $p=1$, and we find 
the convexity condition $a_{33}<0.$   If $s\neq (-)^\sigma$, then (\multip) and (\devide) are rewritten as
$$
\alpha_1^\star=\frac{\pi}{2}
\qquad
{\rm and}
\qquad
\cb=-s\cos\gamma.
\eqn\muldiv
$$
By substituting them to (\symderiv), we find $p=1$ and $s=(-)^\sigma$, contradicting the premise. Thus, there are no solutions in this case.

Now, if we do not assume (\condIIIC), then we have three possibilities. One of them is
$$
\ca=-s\cos\gamma
\qquad
{\rm and}
\qquad
\beta_1^\star=\frac{\pi}{2},
\eqn\rests
$$
which becomes a solution.  By the similar substitution, we acquire $p=1$ and $s=(-)^\sigma$, and the convexity condition is found to be $a_{33}<0$.  No other solutions appear in the remaining possibilities.

\noindent
(ii) For $\phi^\star\neq p\pi$, (\pure) provides the solution for which the convexity condition is given in Table I.

\medskip
\noindent
{\rm (IV-3)  Maximally entangled states:}

Since (\symderiv) becomes (\MESIIi), the solutions and their convexity conditions are the same as those derived in (III-3).

\topinsert
\centerline{
\vbox{
\offinterlineskip
\hrule
\halign{
  $#$ \hfil & $#$ \hfil &  # \hfil & $#$\hfil \cr
 \noalign{\hrule}
\noalign{\smallskip}\cr
 {\rm Conditions}& {\rm Separable} &{\rm Generic} & {\rm Maximally\, entangled} \cr
 \noalign{\smallskip}
\noalign{\hrule}
 \noalign{\smallskip}
 \cases{ a_{33}=0,\cr a_{30}=0}&\forall\alpha_1^\star,\forall\beta_1^\star &  $\forall\alpha_1^\star,\forall\beta_1^\star$ &\forall\alpha_1^\star,\forall\beta_1^\star \cr
 \noalign{\smallskip}
\noalign{\hrule}
 \noalign{\smallskip}
\cases{a_{33}=0,\cr a_{30}\neq0}&\alpha_1^\star=\beta_1^\star=0,\pi& $\alpha_1^\star=\beta_1^\star=0,\pi$&\forall\alpha_1^\star,\forall\beta_1^\star\cr
 \noalign{\smallskip}
\noalign{\hrule}
 \noalign{\smallskip}
\cases{a_{33}\neq0,\cr a_{30}=0}&\left.  \matrix{\sa=\sb=0\cr\ca=\cb=0\cr}\right. &$\left.\matrix{ \sa=\sb=0\cr\cases{\ca=\cb=0,\cr\phi^\star=p\pi\cr}\cr}\right.$& \left.\matrix{\sa=\sb=0\cr\cases{\alpha_1^\star=(-)^p\beta_1^\star+q\pi,\cr\phi^\star=p\pi\cr}\cr}\right.\cr
 \noalign{\smallskip}
\noalign{\hrule}
 \noalign{\smallskip}
\cases{a_{33}<0,\cr a_{30}=(-)^\sigma a_{33}}& \left.  \matrix{\sa=0,\forall\beta_1^\star\cr\sb=0,\forall\alpha_1^\star\cr}\right. &$\left.\matrix{\!\!\!\!\!\!\!\!\!\!\!\!\!\!\!\!\!\!  \sa=\sb=0
\cr
\!\!\!\!\!\!\!\!\!\!\!\!\!\!\!\!\!\!\!\!\! \cases{\ca=\cb\cr=s{r+(-)^{p}\over r-(-)^{p}},\cr\phi^\star=p\pi\cr}\cr\cases{\cos\gamma\ca\cb\cr+\cos\gamma\cr +(-)^\sigma(\ca+\cb)\cr=0,\cr\phi^\star=\pi\cr}\cr}\right.$& \left.\matrix{\sa=\sb=0\cr\cases{\alpha_1^\star=(-)^p\beta_1^\star+q\pi,\cr\phi^\star=p\pi\cr}\cr}\right.\cr
 \noalign{\medskip}
 \noalign{\hrule}
 \noalign{\smallskip}
{\rm Otherwise}& \left.\matrix{\sa=\sb=0\cr\cases{\ca=\cb\cr=(-)^{c+1}s\cr}\cr}\right. &$\left.\matrix{ \sa=\sb=0\cr\cases{\ca=\cb\cr=s{r+(-)^{p}\over r-(-)^{p}},\cr\phi^\star=p\pi\cr}\cr}\right.$& \left.\matrix{\sa=\sb=0\cr\cases{\alpha_1^\star=(-)^p\beta_1^\star+q\pi,\cr\phi^\star=p\pi\cr}\cr}\right.\cr
 \noalign{\smallskip}
 \noalign{\smallskip}
  \noalign{\hrule}
}
\hrule
}
}
\bigskip
\abstract{{\bf Table IV.}  Summary of the complete set of QNE.  Here, $c$, $p$, $q$, and $\sigma$ take values 0 or 1. 
The phase sum $\phi^\star$ is not shown when it is undetermined ({\it i.e.},  any value of $\phi^\star$ is a solution). 
The bottom row with the entry \lq Otherwise\rq\ contains all cases of $a_{33}$ and $a_{30}$ not included in the upper four rows.
}
\bigskip
\endinsert

\vskip 5mm
\centerline{\bf Summary and regrouping of the solutions}
\medskip

The complete set of solutions $(\alpha^\star,\beta^\star)$ obtained above are summarized in Table IV using trigonometric functions for brevity.   These solutions can be regrouped into four distinct types for the convenience of our discussions in the text.

\bigskip
\noindent{\bf Type I}
$$
(\alpha_1^\star,\beta_1^\star)=(k_\alpha,k_\beta)\pi,
\qquad
k_\alpha,k_\beta=0,1.
\eqn\pure
$$
These solutions arise in all of the above four classes.

\bigskip
\noindent{\bf Type II} 
$$
\ca=\cb=s\frac{r+(-)^p}{r-(-)^p}, \qquad \phi^\star=p\pi, \qquad
p=0,1,
\eqn\TypeII
$$
At the separable limit, we have $\ca=\cb=\pm s$. If $s=0$, then $\ca=\cb=0$.

\bigskip
\noindent{\bf Type III} 
$$
\cos\gamma\ca\cb+(-)^\sigma(\ca+\cb)+\cos\gamma=0, 
\qquad
\phi^\star=\pi,
\qquad
s=(-)^\sigma.
\eqn\TypeIII
$$
The separable limit yields $\sa=0$ or $\sb=0$.

\bigskip
\noindent{\bf Type IV}
$$
\alpha_1^\star=(-)^p\beta_1^\star+q\pi,
\qquad
\phi^\star=p\pi.
\eqn\TypeIV
$$
This solution is available only when the joint strategy is maximally entangled, $\gamma = \pi/2$.

\vfill\supereject


\baselineskip= 15.5pt plus 1pt minus 1pt
\parskip=5pt plus 1pt minus 1pt
\tolerance 8000
\vfill\eject\immediate\closeout\reffile
\centerline{{\bf References}}\bigskip\frenchspacing%
\input refs.tmp\vfill\eject\nonfrenchspacing

\bye